\documentclass[twoside,11pt]{article}

\usepackage{multicol}
\usepackage{amsmath}
\usepackage{amsfonts} 
\usepackage{bm}
\usepackage{fancyhdr}
\usepackage{hyperref}

\renewcommand{\sectionmark}[1]%
 {\markboth{\thesection:\ #1}{}}

\numberwithin{equation}{section}

\begin{document}

\title{\bfseries The Spherical Tensor Gradient Operator}

\author{Ernst Joachim Weniger \\
Institut f\"ur Physikalische und Theoretische Chemie \\
Universit\"at Regensburg, D-93040 Regensburg, Germany \\
joachim.weniger@chemie.uni-regensburg.de}

\date{To Appear in the Josef Paldus Birthday lssue of \\
Collection of Czechoslovak Chemical Communications \\[1.5\jot]
Date of Submision: 3 February 2005 \\
Date of Acceptance: 15 April 2005 \\
Final Submission: 6 May 2005}

\maketitle

\noindent
\textbf{Dedicated to Professor Josef Paldus on the occasion of his 70th
  birthday}

\bigskip



\bigskip

\typeout{==> Abstract}
\begin{abstract}
\noindent
The spherical tensor gradient operator ${\mathcal{Y}}_{\ell}^{m}
(\nabla)$, which is obtained by replacing the Cartesian components of
$\bm{r}$ by the Cartesian components of $\nabla$ in the regular
solid harmonic ${\mathcal{Y}}_{\ell}^{m} (\bm{r})$, is an irreducible
spherical tensor of rank $\ell$. Accordingly, its application to a
scalar function produces an irreducible spherical tensor of rank $\ell$.
Thus, it is in principle sufficient to consider only multicenter
integrals of scalar functions: Higher angular momentum states can be
generated by differentiation with respect to the nuclear coordinates.
Many of the properties of ${\mathcal{Y}}_{\ell}^{m} (\nabla)$ can be
understood easily with the help of an old theorem on differentiation by
Hobson [Proc.\ London Math.\ Soc.\ {\bf 24}, 54 - 67 (1892)]. It follows
from Hobson's theorem that some scalar functions of considerable
relevance as for example the Coulomb potential, Gaussian functions, or
reduced Bessel functions produce particularly compact results if
${\mathcal{Y}}_{\ell}^{m} (\nabla)$ is applied to them. Fourier
transformation is very helpful to understand the properties of
${\mathcal{Y}}_{\ell}^{m} (\nabla)$ since it produces
${\mathcal{Y}}_{\ell}^{m} (-\mathrm{i} \bm{p})$. It is also possible to
apply ${\mathcal{Y}}_{\ell}^{m} (\nabla)$ to generalized functions,
yielding for instance the spherical delta function $\delta_{\ell}^{m}
(\bm{r})$. The differential operator ${\mathcal{Y}}_{\ell}^{m} (\nabla)$
can also be used for the derivation of pointwise convergent addition
theorems. The feasibility of this approach is demonstrated by deriving
the addition theorem of $r^{\nu} {\mathcal{Y}_{\ell}^{m}} (\bm{r})$ with
$\nu \in \mathbb{R}$.
\end{abstract}

\bigskip

\tableofcontents

\newpage

\typeout{==> Section: Introduction}
\section{Introduction}
\label{Sec:Intro}

In the early 17th century, scientific methodology advanced significantly
by what Kline \cite[Chapter 16]{Kline/1972} called the mathematization
of science. Roughly at the same time, the foundations for the later
development of calculus were laid (see for example \cite[Chapter
17]{Kline/1972}). Differential and integral calculus greatly extended
the arsenal of mathematical techniques that could be used for a
description and analysis of scientific phenomena. Ultimately, this
development led to a period of unity between mathematics and sciences
which lasted approximately to the end of the 19th or to the beginning of
the 20th century. In that period, it was frequently not possible to
decide whether somebody was predominantly a mathematician or
predominantly a scientist, and mathematics and the sciences enriched
each other greatly by cross fertilization.

A striking example for this unity between mathematics and the sciences
is provided by Peter Debye who had been a student of Sommerfeld. Among
chemists, Debye is best known for his work on electrolyte solutions,
which earned him the Nobel Prize in Chemistry in 1936. It is, however,
not nearly so well known that he was also an excellent mathematician who
made significant contribution to the theory of Bessel and Hankel
functions (see for example \cite[Chapter VIII]{Watson/1966}).

Since the times of Debye, the amount of collective knowledge has
increased tremendously. Accordingly, contemporary research is
predominantly done by specialists, who know almost everything about almost
nothing, whereas generalists like Debye, who had done excellent research
in mathematics, physics, and physical chemistry, have become exceedingly
rare. In particular, there is a widening gap between mathematicians and
those that apply mathematics. In the past, mathematicians had developed
new analytical or numerical techniques that greatly helped scientists
and engineers to do research in their disciplines. In return, open
scientific or engineering problems had always provided a valuable source
of inspiration for mathematicians. Unfortunately, this is now longer
true. Due to increased specialization, communication between mathematics
and those disciplines, that use mathematics as their main language, has
deteriorated. This is certainly a deplorable development because cross
fertilizations typically happens at the interfaces of different
disciplines. 

Fortunately, encouraging counterexamples still exist: For example,
Professor Josef Paldus started his career at the Heyrovsk\'{y} Institute
in Prague doing experimental and theoretical work in electrochemistry,
and he later also worked as an experimental spectroscopist. In addition,
he soon ventured into the newly emerging field of quantum chemistry. In
1968, he came as Brezhnev's gift to the Department of Applied
Mathematics of the University of Waterloo. There, his predominant
research interest has been the treatment of electronic correlation,
emphasizing highly sophisticated mathematical techniques as for example
the representation theory of Lie groups. It was this strong mathematical
orientation of his research which earned him -- in addition to numerous
other honors -- the Fellowship of the prestigious Fields Institute for
Research in Mathematical Sciences.

Quite in the spirit of the highly interdisciplinary research of
Professor Josef Paldus, I want to discuss in this article a topic that
is located somewhere at the interface between mathematics and atomic and
molecular electronic structure theory.

Since space is isotropic, free atoms are spherically symmetric and
angular momentum is conserved. Thus, spherical polar coordinates and the
machinery of angular momentum theory lead to significant computational
and conceptual simplifications in atomic structure calculations. It is
probably only a slight exaggeration to claim that atomic electronic
structure theory is essentially an application of angular momentum
theory.

In the case of molecules, the benefits of spherical coordinates and
angular momentum theory are not so obvious. In general, molecules
possess no spatial symmetry at all, and if they have one, it is a lower
symmetry than spherical symmetry. Thus, neither spherical polar
coordinates nor angular momentum theory lead to a such spectacular
reduction of computational complexity as they do in the case of atoms.

However, chemists have found it helpful to approach molecular electronic
structure theory from the perspective of atoms. For example, the most
successful computational scheme, the so-called LCAO-MO approach, is
based on the tacit assumption that the parentage of atoms facilitates
our attempts of understanding the electronic structure of molecules.
Thus, in molecular electronic structure calculations it is common to use
angular momentum theory at least locally, i.e., with respect to the
atomic centers. In this way, at least some of the formal advantages of
angular momentum theory can be retained.

Nevertheless, the use of angular momentum theory for systems lacking
rotational symmetry causes new mathematical and computational
challenges. In the case of atoms, only relative small angular momentum
quantum numbers occur, and the coupling of irreducible spherical tensors
produces finite expressions consisting of a few terms only. In the case
of molecules, the situation is in general much more complicated: We have
to deal with infinite series expansions over angular momentum states
that do not necessarily converge rapidly. Thus, we must be able to
compute the typical quantities of angular momentum theory both
efficiently and reliably even for very large angular momentum quantum
numbers. Quite often, this is not so easy.

In molecular electronic structure calculations with analytic basis
functions centered at the atomic nuclei, the basic quantities are matrix
elements, i.e., essentially multicenter integrals whose evaluation may
be very difficult. It is an undeniable empirical fact that it is often
comparatively easy to obtain explicit analytical expressions for
multicenter integrals over functions that are scalars or irreducible
spherical tensors of rank zero with respect to their local (atomic)
coordinate systems. If, however, the functions occurring in the
multicenter integral are irreducible spherical tensors of higher ranks,
one can easily get lost in an algebraic jungle and the derivation of
explicit expressions can become extremely difficult or even impossible.

It is another empirical fact that it is usually much easier to
differentiate than to integrate. Accordingly, it is an obvious idea to
try to generate an explicit expression for a multicenter integral over
nonscalar functions by differentiating the simpler expression for the
corresponding integral over scalar functions (preferably the simplest
scalar functions) with respect to scaling parameters and/or nuclear
coordinates (compare also \cite[Section IV]{Weniger/Steinborn/1983a}).
The use of generating differential operators does not necessarily
produce closed form expression that hold for arbitrary quantum numbers.
Instead, it may be necessary to derive a new expression for each set of
quantum numbers. Obviously, a multitude of special formulas is not
nearly as convenient as a neat general explicit expression, but powerful
computer algebra systems like Maple or Mathematica with the ability of
automatically generating FORTRAN or C code help to make such an approach
feasible (see for example \cite{Bracken/Bartlett/1997}).

It is relatively easy to generate multicenter integrals of higher scalar
functions by differentiating the most simple scalar functions with
respect to their scaling parameters. In the case of a $1s$ Slater-type
or Gaussian function, we can construct higher scalar functions easily by
repeatedly using the relationships $\partial \exp (-\alpha r)/\partial
\alpha = -r \exp (-\alpha r)$ or $\partial \exp (-\alpha r^2)/\partial
\alpha = -r^2 \exp (-\alpha r)$.

The generation of anisotropic functions, that are irreducible spherical
tensors of rank $\ell$, from scalar functions is less straightforward:
It follows from Hobson's theorem \cite{Hobson/1892}, which is discussed
in Section \ref{Sec:HobsonDiffTheor}, that we have to apply the
spherical tensor gradient operator ${\mathcal{Y}}_{\ell}^{m} (\nabla)$
to scalar functions. This differential operator is obtained by
replacing in the regular solid harmonic ${\mathcal{Y}}_{\ell}^{m}
(\bm{r}) = r^{\ell} Y_{\ell}^{m} (\theta, \phi)$ the Cartesian
components of $\bm{r} = (x, y, z)$ by the Cartesian components of
$\nabla = (\partial/\partial x, \partial/\partial y, \partial/\partial
z)$.

I came across the differential operator ${\mathcal{Y}}_{\ell}^{m}
(\nabla)$ during my PhD thesis \cite{Weniger/1982} in which I worked on
multicenter integrals of reduced Bessel functions and their anisotropic
generalizations, the so-called $B$ functions. These functions, which are
discussed in Section \ref{Sec:RBF}, are a special class of exponentially
decaying functions with some very useful properties. When I derived the
remarkably compact Fourier transform (\ref{FT_B_Fun}) of a $B$ function
\cite[Eq.\ (7.1-6) on p.\ 160]{Weniger/1982}, I noticed that the Fourier
integral representation (\ref{InvFT_B_Fun}) of an anisotropic $B$
function $B_{n,\ell}^{m} (\alpha, \bm{r})$ differs from that of the
scalar $B$ function $B_{n+\ell, 0}^{m} (\alpha, \bm{r})$ only by an
additional regular solid harmonic $\mathcal{Y}_{\ell}^{m} (-\mathrm{i}
\bm{p})$. Because of (\ref{YlmNabla_PlaneWave}),
$\mathcal{Y}_{\ell}^{m} (\mathrm{i} \bm{p})$ can be interpreted to be
the Fourier transform of the differential operator
$\mathcal{Y}_{\ell}^{m} (\nabla)$, and I immediately deduced that the
anisotropic $B$ function $B_{n,\ell}^{m} (\alpha, \bm{r})$ can according
to (\ref{STGO_Bn00}) be generated by applying $\mathcal{Y}_{\ell}^{m}
(\nabla)$ to the scalar $B$ function $B_{n+\ell, 0}^{m} (\alpha,
\bm{r})$ \cite[Eq.\ (7.1-10) on p.\ 161]{Weniger/1982}.

Of course, this observation aroused my interest: I wanted to know
whether this result can also be proved directly without the help of
Fourier transformation, and I also wanted to know whether the
differential operator $\mathcal{Y}_{\ell}^{m} (\nabla)$ is a useful
analytical tool in different contexts. After the completion of my PhD
thesis, I studied the spherical tensor gradient operator more seriously.
I soon learned that I was not the only one and in particular not the
first one who had studied and used this differential operator: The first
article dealing with the spherical tensor gradient operator, which I am
aware of, was published by Hobson \cite{Hobson/1892} in 1892. I also
noticed that $\mathcal{Y}_{\ell}^{m} (\nabla)$ is a highly useful
analytical tool for a wide range of problems. For example, I found
recent articles which describe successful applications of
$\mathcal{Y}_{\ell}^{m} (\nabla)$ in scattering
\cite{Bracher/Kramer/Kleber/2003} or in solid state
theory \cite{Bott/Methfessel/Krabs/Schmidt/1998,%
Methfessel/VanSchilfgaarde/Casali/1999}.

From the perspective of quantum chemistry, it is probably more
interesting that $\mathcal{Y}_{\ell}^{m} (\nabla)$ can be extremely
useful in the context of molecular multicenter integrals of
exponentially decaying functions, as shown in articles by Grotendorst
and Steinborn 
\cite{Grotendorst/Steinborn/1985,Grotendorst/Steinborn/1988}, Niukkanen
\cite{Niukkanen/1983a,Niukkanen/1984b,Niukkanen/1985b}, Novosadov
\cite{Novosadov/1983,Novosadov/1988,Novosadov/2001a,Novosadov/2002a,%
Novosadov/2002b,Novosadov/2003}, Tai \cite{Tai/1994}, and in my own
research \cite{Steinborn/Weniger/1992,Weniger/2000a,Weniger/2002,%
Weniger/Grotendorst/Steinborn/1986b,Weniger/Steinborn/1983a,%
Weniger/Steinborn/1983c,Weniger/Steinborn/1985,Weniger/Steinborn/1989b}.

The spherical tensor gradient operator $\mathcal{Y}_{\ell}^{m} (\nabla)$
is -- as discussed in more details in Section \ref{Sec:HobsonDiffTheor}
-- also extremely useful in the context of multicenter integrals of
spherical Gaussian functions. For example, Dunlap reformulated in a
series of recent articles
\cite{Dunlap/1990,Dunlap/2001,Dunlap/2002,Dunlap/2003,Dunlap/2005} the
theory of multicenter integrals of anisotropic spherical Gaussians by
systematically applying $\mathcal{Y}_{\ell}^{m} (\nabla)$ to the
corresponding integrals of $1s$ Gaussians. Many more references on
multicenter integrals of spherical Gaussians can be found in Section
\ref{Sec:HobsonDiffTheor}.

These examples should suffice to show that $\mathcal{Y}_{\ell}^{m}
(\nabla)$ is indeed a very useful analytical tool that has been applied
successfully to a wide range of problems. Nevertheless, a reasonably
comprehensive discussion the mathematical properties of the spherical
tensor gradient operator from the perspective of a potential future user
seems to be missing. This is what I intend to provide with this article.

In Section \ref{Sec:Ylm_YlmNabla}, the spherical tensor gradient
operator is introduced and some general features -- in particular those
based on its tensorial nature -- are discussed. 

Section \ref{Sec:HobsonDiffTheor} treats Hobson's theorem on
differentiation, by means of which many properties of
$\mathcal{Y}_{\ell}^{m} (\nabla)$ can be derived and understood easily.

Fourier transformation, which is discussed in Section
\ref{Sec:FourierTr}, maps the gradient operator $\nabla$ to $\mathrm{i}
\bm{p}$, or equivalently $\mathcal{Y}_{\ell}^{m} (\nabla)$ to
$\mathcal{Y}_{\ell}^{m} (\mathrm{i} \bm{p})$. Thus, in momentum space,
we only have to study the comparatively simple multiplicative operator
$\mathcal{Y}_{\ell}^{m} (\mathrm{i} \bm{p})$ whose properties are very
well understood. This greatly facilitates a theoretical analysis.

In Section \ref{Sec:RBF}, the so-called reduced Bessel functions and
their anisotropic generalizations, the so-called $B$ functions, are
discussed. These functions play a special role among exponentially
decaying functions because of their exceptionally simple Fourier
transform and because of the ease, with which the spherical tensor
gradient operator can be applied. 

Classically, the domain of the spherical tensor gradient operator
consists of the differentiable functions $f: \mathbb{R}^3 \to
\mathbb{C}$, but it makes sense to apply it also to generalized
functions. Thus, in Section \ref{Sec:SpherDeltaFun} the spherical delta
function and related objects -- for example distributional $B$ functions
-- are treated. 

In Section \ref{Sec:AdditionThm}, the derivation of addition theorems of
essentially arbitrary irreducible spherical tensors with the help of the
spherical tensor gradient operator is discussed. In addition, the
addition theorem of $r^{\nu} {\mathcal{Y}_{\ell}^{m}} (\bm{r})$ with
$\nu \in \mathbb{R}$ is constructed in order to demonstrate the
feasibility of the whole approach. 

This article is concluded by a short summary in Section
\ref{Sec:SummConcl}. Finally, there are three Appendices: The
terminology used in this article is introduced in Appendix
\ref{App:Terminolgy}, the for our purposes most relevant properties of
the spherical harmonics are reviewed in Appendix \ref{App:SpherHar}, and
Gaunt coefficients are discussed in Appendix \ref{App:Gaunt}.

\typeout{==> Section: The Spherical Tensor Gradient Operator}
\section{The Spherical Tensor Gradient Operator}
\label{Sec:Ylm_YlmNabla}

The spherical tensor gradient operator ${\mathcal{Y}}_{\ell}^{m}
(\nabla)$ can be introduced via the explicit expression
(\ref{YlmHomPol}) for the regular solid harmonic
${\mathcal{Y}}_{\ell}^{m} (\bm{r})$. This explicit expression holds for
the Cartesian components of essentially arbitrary three-dimensional
vectors. Thus, we obtain the differential operator
${\mathcal{Y}}_{\ell}^{m} (\nabla)$ if we replace in
(\ref{HomogLaplaceEqn}) the Cartesian components of $\bm{r} = (x, y, z)$
by the Cartesian components of $\nabla = (\partial/\partial x,
\partial/\partial y, \partial/\partial z)$:
{\allowdisplaybreaks
\begin{align}
  \label{Def:YlmNabla}
  \mathcal{Y}_{\ell}^{m} (\nabla) & \; = \; \left[
    \frac{2\ell+1}{4\pi} (\ell+m)!(\ell-m)! \right]^{1/2}
  \notag \\[1.5\jot]
  & \qquad \times \, \sum_{k \ge 0} \, \frac
  {\left(-\frac{\partial}{\partial x} -
      \mathrm{i}\frac{\partial}{\partial y}\right)^{m+k} \,
    \left(\frac{\partial}{\partial x} -
      \mathrm{i}\frac{\partial}{\partial y}\right)^{k} \, \left(
      \frac{\partial}{\partial z} \right)^{\ell-m-2k}} {2^{m+2k} (m+k)!
    k! (\ell-m-2k)!} \, .
\end{align}
}

A new differential operators is not necessarily a useful thing, let
alone a major achievement. In atomic and molecular calculations, we are
interested in functions that are in defined in terms of spherical polar
coordinates and that can be expressed as a radial part multiplied by a
spherical harmonic:
\begin{equation}
  \label{Def_IrrSphericalTensor}
F_{\ell}^{m} (\bm{r}) \; = \; 
f_{\ell} (r) \, Y_{\ell}^{m} (\bm{r}/r) \, .
\end{equation}
The straightforward differentiations of such an irreducible tensor of
rank $\ell$ with respect to the Cartesian components $x$, $y$, and $z$
of $\bm{r}$ would in general lead to extremely messy expressions.
Because of the convenient orthonormality properties of the spherical
harmonics it would be advantageous if the angular part of such a product
could be expressed in terms of spherical harmonics. Of course, it should
be possible to do the necessary algebra, but in particular for large
angular momentum quantum numbers we would be confronted with nontrivial
technical problems.

Thus, we arrive at the paradoxical statement that the differential
operator ${\mathcal{Y}}_{\ell}^{m} (\nabla)$ is practically useful only
if it is not necessary to do differentiations with respect to $x$, $y$,
and $z$ via the defining explicit expression (\ref{Def:YlmNabla}).
Fortunately, these differentiations can be avoided since
${\mathcal{Y}}_{\ell}^{m} (\nabla)$ is just like the corresponding
regular solid harmonic ${\mathcal{Y}}_{\ell}^{m} (\bm{r})$ an
irreducible spherical tensor of rank $\ell$ (compare \cite[p.\ 
312]{Biedenharn/Louck/1981a}). Consequently, matrix elements involving
${\mathcal{Y}}_{\ell}^{m} (\nabla)$ and other irreducible spherical
tensors can be handled via the powerful machinery of angular momentum
coupling.

As discussed in more details later, the product
${\mathcal{Y}}_{\ell_1}^{m_1} (\nabla) F_{\ell_2}^{m_2} (\bm{r})$
can be expressed as a finite linear combination of Gaunt coefficients
defined in (\ref{Def_Gaunt}), radial functions $\gamma _{\ell_1
  \ell_2}^{\ell} (r)$, and spherical harmonics \cite[Eq.\ 
(4.7)]{Weniger/Steinborn/1983c}:
\begin{align}
  \label{YlmNab2Flm_GenStruc}
  & {\mathcal{Y}}_{\ell_1}^{m_1} (\nabla) \, F_{\ell_2}^{m_2}
  (\bm{r})
  \notag \\
  & \qquad \; = \;
  \sum_{\ell=\ell_{\mathrm{min}}}^{\ell_{\mathrm{max}}} \! {}^{(2)}
  \, \langle \ell m_1+m_2 \vert \ell_1 m_1 \vert \ell_2 m_2 \rangle \,
  \gamma _{\ell_1 \ell_2}^{\ell} (r) \, Y_{\ell}^{m_1+m_2} (\bm{r}/r) 
  \, .
\end{align}
The summation limits $\ell_{\mathrm{min}}$ and $\ell_{\mathrm{max}}$ are
given by (\ref{SumLims}).

It is possible to derive alternatives to (\ref{YlmNab2Flm_GenStruc})
which also take into account the tensorial nature of both
${\mathcal{Y}}_{\ell_1}^{m_1} (\nabla)$ and $F_{\ell_2}^{m_2}
(\bm{r})$. For example, Bayman \cite{Bayman/1978} derived an equivalent
expression involving Clebsch-Gordan coefficients instead of Gaunt
integrals as in (\ref{YlmNab2Flm_GenStruc}). However, it was argued in
\cite[Section III]{Weniger/Steinborn/1983c} that $\nabla$ behaves
like $\bm{r}$ under reflections. Accordingly, parity is conserved, and
the replacement of Clebsch-Gordan coefficients by Gaunt integrals
according to (\ref{Gaunt_ClebschGordan}) leads to a considerable formal
simplification.

As discussed in more details in Section \ref{Sec:FourierTr}, the
functions $\gamma _{\ell_1 \ell_2}^{\ell} (r)$ in
(\ref{YlmNab2Flm_GenStruc}) can be obtained by differentiating the
radial part $f_{\ell_2} (r)$ of the spherical tensor $F_{\ell_2}^{m_2}
(\bm{r})$ with respect to $r$ (compare the discussion in \cite[Sections
III and IV]{Weniger/Steinborn/1983c}). Thus, the systematic exploitation
of the tensorial nature of ${\mathcal{Y}}_{\ell_1}^{m_1} (\nabla)$
makes it possible to replace potentially troublesome differentiations
with respect to the three Cartesian components $x$, $y$, and $z$ by
comparatively benign differentiations with respect to the radial
variable $r$. This is a very important technical aspect since it greatly
facilitates the application of the spherical tensor gradient operator to
spherical tensors of the type of (\ref{Def_IrrSphericalTensor}).

Other expressions for the product ${\mathcal{Y}}_{\ell_1}^{m_1}
(\nabla) F_{\ell_2}^{m_2} (\bm{r})$ can be found in articles by
Santos \cite{Santos/1973}, Stuart \cite{Stuart/1981}, Niukkanen
\cite{Niukkanen/1983a}, and Rashid \cite{Rashid/1986}.

\typeout{==> Section: Hobson's Theorem on Differentiation}
\section{Hobson's Theorem on Differentiation}
\label{Sec:HobsonDiffTheor}

The first article on the spherical tensor gradient operator
${\mathcal{Y}}_{\ell}^{m} (\nabla)$, which I am aware of, is due to
Hobson \cite[p.\ 67]{Hobson/1892} who derived in 1892 a very
consequential theorem on the differentiation of functions $f:
\mathbb{R}^n \to \mathbb{C}$. This theorem is also discussed in Hobson's
book \cite[pp.\ 124 - 129]{Hobson/1965} that was first published in
1931.
 
In contemporary articles dealing with the spherical tensor gradient
operator, Hobson's theorem -- Eq.\ (\ref{Hobson_Thm_HomDiffOp}) below --
is in most cases completely ignored, but in the article by Bott,
Methfessel, Krabs, and Schmidt \cite[Section
IV]{Bott/Methfessel/Krabs/Schmidt/1998}, the simplified form
(\ref{Hobson_Simpli_Thm}) of Hobson's theorem was even rederived in a
less general form. This is an undeserved neglect. Firstly, many
properties of the spherical tensor gradient operator
${\mathcal{Y}}_{\ell}^{m} (\nabla)$ can be deduced and understood easily
via Hobson's theorem. Secondly, it should be possible to generalize the
approach described in this article, which is based on the
three-dimensional regular solid harmonics ${\mathcal{Y}}_{\ell}^{m}
(\bm{r})$, to differential operators based on $n$-dimensional
hyperspherical harmonics. A problem of obvious physical relevance would
be the construction of differential operators associated to the
four-dimensional hyperspherical harmonics, which are for instance
discussed in books by Avery \cite{Avery/1989,Avery/2000} and Judd
\cite[Sections 2 and 3 and Appendix 2]{Judd/1975} or also in
\cite[Section VI]{Weniger/1985}.

In three-dimensional form -- which is all we need here -- Hobson's
theorem can be formulated as follows (compare also \cite[Eq.\ 
(4.13)]{Weniger/Steinborn/1983a}):

{\noindent \sl Let $f_{n} (x, y, z)$ be a homogeneous polynomial of
  degree $n \in \mathbb{N}$ in the variables $x, y, z$, and let $F$ be
  any (differentiable) function that depends only on $r^2 = x^2 + y^2 +
  z^2$. Then, the application of the differential operator $f_{n}
  (\partial/\partial x, \partial/\partial y, \partial/\partial z)$ to
  $F$ can be expressed in closed form according to}
\begin{align}
  \label{Hobson_Thm_HomDiffOp}
  & f_{n} \left( \frac{\partial}{\partial x}, \frac{\partial}{\partial
      y}, \frac{\partial}{\partial z} \right) F (r^2)
  \notag \\
  & \qquad \; = \; \sum_{\nu=0}^{n} \, \frac{2^{n-2\nu}}{\nu!} \, \left[
    \left( \frac{\mathrm{d}}{\mathrm{d} r^2} \right)^{n-\nu} \, F (r^2)
  \right] \, \nabla^{2\nu} \, f_{n} (x, y, z) \, .
\end{align}
Let us now assume that $f_n$ is not only a homogeneous polynomial of
degree $n$, but also a solution of the three-dimensional homogeneous
Laplace equation (\ref{HomogLaplaceEqn}). Then,
(\ref{Hobson_Thm_HomDiffOp}) simplifies considerably because in the sum
on the right-hand side only the power $\nabla^{2\nu}$ with $\nu =
0$ produces a nonzero result:
\begin{equation}
  \label{Hobson_Simpli_Thm}
f_n \left( \frac{\partial}{\partial x}, \frac{\partial}{\partial y},
\frac{\partial}{\partial z} \right) F (r^2) \; = \; 2^{n} \, \left[
\left( \frac{\mathrm{d}}{\mathrm{d} r^2} \right)^{n} \, F (r^2) 
\right] \, f_n (x, y, z) \, .
\end{equation}

As discussed in Appendix \ref{App:SpherHar}, the regular solid harmonic
$\mathcal{Y}_{\ell}^{m} (\bm{r})$ is a polynomial solution of the
homogeneous Laplace equation (\ref{HomogLaplaceEqn}) of degree $\ell$.
Accordingly, Hobson's theorem applies in its simplified form
(\ref{Hobson_Simpli_Thm}), and we obtain (compare \cite[Eq.\ 
(3.11)]{Weniger/Steinborn/1983a}):
\begin{equation}
  \label{YlmNabla_F}
{\mathcal{Y}}_{\ell}^{m} (\nabla) \, F (r^2) \; = \; 2^{\ell} \, 
\biggl[ \left( \frac{\mathrm{d}}{\mathrm{d} r^2} \right)^{\ell} \, 
F (r^2) \biggr] \, {\mathcal{Y}}_{\ell}^{m} (\bm{r}) \, .
\end{equation}
If we now set $F (r^2) = \varphi (r)$ and use $\mathrm{d}/\mathrm{d} r^2
= 1/(2r) \, \mathrm{d}/\mathrm{d} r$, we obtain (compare \cite[Eq.\ 
(3.12)]{Weniger/Steinborn/1983a}):
\begin{equation}
  \label{YlmNabla_phi}
{\mathcal{Y}}_{\ell}^{m} (\nabla) \, \varphi (r) \; = \; \biggl[ 
\left( \frac{1}{r} \, \frac{\mathrm{d}}{\mathrm{d} r} \right)^{\ell} 
\, \varphi (r) \biggr] \, {\mathcal{Y}}_{\ell}^{m} (\bm{r}) \, .
\end{equation}
It follows from either (\ref{YlmNabla_F}) or (\ref{YlmNabla_phi}) that
${\mathcal{Y}}_{\ell}^{m} (\nabla)$ can be viewed as a generating
differential operator which transforms a scalar function -- an
irreducible spherical tensor of rank $0$ -- to an irreducible spherical
tensor of rank $\ell$.

Next, let us apply the spherical tensor gradient operator
${\mathcal{Y}}_{\ell_1}^{m_1} (\nabla)$ to irreducible spherical
tensors $F_{\ell_2}^{m_2}$ of the type of (\ref{Def_IrrSphericalTensor})
with nonzero rank $\ell_2$. To simplify things, let us for the moment
assume that such a spherical tensor $F_{\ell_2}^{m_2} (\bm{r})$
satisfies a relation of the kind of (\ref{YlmNabla_phi}), i.e., it can
be generated by applying ${\mathcal{Y}}_{\ell_2}^{m_2} (\nabla)$ to
a suitable scalar function $\Phi_{\ell_2} (r)$:
\begin{equation}
  \label{Def_Phi_l2m2}
F_{\ell_2}^{m_2} (\bm{r}) \; = \; 
{\mathcal{Y}}_{\ell_2}^{m_2} (\nabla) \, \Phi_{\ell_2} (r) \, .
\end{equation}
In view of ${\mathcal{Y}}_{\ell_1}^{m_1} (\nabla) F_{\ell_2}^{m_2}
(\bm{r}) = {\mathcal{Y}}_{\ell_1}^{m_1} (\nabla)
{\mathcal{Y}}_{\ell_2}^{m_2} (\nabla) \, \Phi_{\ell_2} (r)$, we
need an explicit expression of manageable complexity for the product
${\mathcal{Y}}_{\ell_1}^{m_1} (\nabla) {\mathcal{Y}}_{\ell_2}^{m_2}
(\nabla)$. This can be accomplished easily. If we multiply the
linearization formula (\ref{Ylm_lin}) of the surface harmonics by
$r^{\ell_1+\ell_2}$, we obtain the linearization formula of the regular
solid harmonics:
\begin{align}
  \label{RegSolYlm_lin}
  & \mathcal{Y}_{\ell_1}^{m_1} (\bm{r}) \, \mathcal{Y}_{\ell_2}^{m_2}
  (\bm{r})
  \notag \\
  & \qquad \; = \;
  \sum_{\ell=\ell_{\mathrm{min}}}^{\ell_{\mathrm{max}}} \! {}^{(2)}
  \, \langle \ell m_1+m_2 \vert \ell_1 m_1 \vert \ell_2 m_2 \rangle \,
  r^{2 \Delta \ell} \, \mathcal{Y}_{\ell}^{m_1+m_2} (\bm{r}) \, .
\end{align}
It follows at once from the summation limits (\ref{SumLims}) that the
abbreviation $\Delta \ell$ defined by (\ref{Def_Del_l}) is either a
positive integer or zero. Accordingly, (\ref{RegSolYlm_lin}) can also be
interpreted as a linearization formula for certain polynomials in the
Cartesian components of an essentially arbitrary three-dimensional
vector $\bm{r}$. Thus, (\ref{RegSolYlm_lin}) also holds for $\bm{r} =
\nabla$, and we obtain (see for example \cite[Eq.\ 
(15)]{Niukkanen/1983a}, \cite[Eq.\ (65)]{Novosadov/1983}, or \cite[Eq.\ 
(4.24)]{Weniger/Steinborn/1983a}):
\begin{align}
  \label{YlmNabla_lin}
  & \mathcal{Y}_{\ell_1}^{m_1} (\nabla) \,
  \mathcal{Y}_{\ell_2}^{m_2} (\nabla)
  \notag \\
  & \qquad \; = \;
  \sum_{\ell=\ell_{\mathrm{min}}}^{\ell_{\mathrm{max}}} \! {}^{(2)}
  \, \langle \ell m_1+m_2 \vert \ell_1 m_1 \vert \ell_2 m_2 \rangle \,
  \nabla^{2 \Delta \ell} \, \mathcal{Y}_{\ell}^{m_1+m_2}
  (\nabla) \, .
\end{align}
If we now combine (\ref{Def_Phi_l2m2}) and (\ref{YlmNabla_lin}), we
obtain an expression which is obviously of the form of
(\ref{YlmNab2Flm_GenStruc}) (see for example \cite[Eq.\ 
(3.9)]{Weniger/Steinborn/1985}):
\begin{align}
  \label{YlmNabla_PHIlm}
  {\mathcal{Y}}_{\ell_1}^{m_1} (\nabla) \, F_{\ell_2}^{m_2}
  (\bm{r}) & \; = \;
  \sum_{\ell=\ell_{\mathrm{min}}}^{\ell_{\mathrm{max}}} \! {}^{(2)}
  \, \langle \ell m_1+m_2 \vert \ell_1 m_1 \vert \ell_2 m_2 \rangle
  \notag \\
  & \qquad \times \, \nabla^{2 \Delta \ell} \, \biggl[ \left(
    \frac{1}{r} \, \frac{\mathrm{d}}{\mathrm{d} r} \right)^{\ell} \,
  \Phi_{\ell_2} (r) \biggr] \, {\mathcal{Y}}_{\ell}^{m_1+m_2} (\bm{r})
  \, .
\end{align}

There are some radially symmetric functions of considerable physical
relevance which produce results of remarkable simplicity if
${\mathcal{Y}}_{\ell}^{m} (\nabla)$ is applied to them via
(\ref{YlmNabla_phi}). The classic example is the Coulomb potential
$1/r$. Hobson \cite{Hobson/1892} showed that the irregular solid
harmonic $\mathcal{Z}_{\ell}^{m} (\bm{r})$ defined in
(\ref{Def_IrregSolHar}) is generated by applying
${\mathcal{Y}}_{\ell}^{m} (\nabla)$ to $1/r$ (further details can
be found in Hobson's book \cite[pp.\ 124 - 129]{Hobson/1965}). In modern
notation, Hobson's result can be expressed as follows (see for example
\cite[Eq.\ (4.16)]{Weniger/Steinborn/1983a}):
\begin{equation}
   \label{YllNabla_Coul}
\mathcal{Z}_{\ell}^{m} (\bm{r}) \; = \;
\frac{(-1)^{\ell}}{(2 \ell -  1)!!}  
\, \mathcal{Y}_{\ell}^{m} (\nabla) \, \frac{1}{r} \, .
\end{equation}
It is possible to derive this relationships differently (see for example
\cite[Chapter 6.18]{Biedenharn/Louck/1981a}), but its derivation via
(\ref{YlmNabla_phi}) is in my opinion much more straightforward and
transparent. 

It is also quite easy to obtain an explicit expression for the product
$\mathcal{Y}_{\ell_1}^{m_1} (\nabla) \mathcal{Z}_{\ell_2}^{m_2}
(\bm{r})$. With the help of (\ref{YlmNabla_PHIlm}) and
(\ref{YllNabla_Coul}), we immediately obtain the following compact
expression involving Gaunt coefficients, Pochhammer symbols, and
irregular solid harmonics \cite[Eq.\ (4.7)]{Weniger/Steinborn/1985}:
\begin{align}
  \label{YlmNabla_Zlm}
  \mathcal{Y}_{\ell_1}^{m_1} (\nabla) \, \mathcal{Z}_{\ell_2}^{m_2}
  (\bm{r}) & \; = \; \langle \ell_1 + \ell_2 \, m_1+m_2 \vert \ell_1 m_1
  \vert \ell_2 m_2 \rangle
  \notag \\
  & \qquad \times (-2)^{\ell_1} \, \frac{(1/2)_{\ell_1 +
      \ell_2}}{(1/2)_{\ell_2}} \, \mathcal{Z}_{\ell_1+\ell_2}^{m_1+m_2}
  (\mathbf{r}) \, .
\end{align}
Alternative derivations of this result are discussed in \cite[Section
7]{Weniger/2000a}.

In the vast majority of all LCAO-MO electronic structure calculations,
Gaussian functions are used as basis functions. Traditionally, Cartesian
Gaussian functions $x^i y^j z^k \exp (-\alpha r^2)$ with $i, j, k \in
\mathbb{N}_0$ and $\alpha \in \mathbb{R}_{+}$ have dominated, which
indicates that the spherical tensor gradient operator should be
irrelevant for calculations with Gaussian basis functions. However, in
recent years research on molecular integrals of Gaussians has
increasingly emphasized Gaussian functions whose angular parts are
spherical harmonics. For these functions, ${\mathcal{Y}}_{\ell}^{m}
(\nabla)$ is again an extremely useful mathematical tool (see for
example the articles by Arakane and Matsuoka
\cite{Arakane/Matsuoka/1998}, Chow Chiu and Moharerrzadeh
\cite{ChowChiu/Moharerrzadeh/1998,ChowChiu/Moharerrzadeh/1999,%
ChowChiu/Moharerrzadeh/2001}, Dunlap
\cite{Dunlap/1990,Dunlap/2001,Dunlap/2002,Dunlap/2003,Dunlap/2005},
Fieck \cite{Fieck/1979,Fieck/1980}, Fortunelli and Carrravetta
\cite{Fortunelli/Carrravetta/1992}, Fortunelli and Salvetti
\cite{Fortunelli/Salvetti/1993}, Fujimura and Matsuoka
\cite{Fujimura/Matsuoka/1992}, Hu, Staufer, Birkenheuer, Igoshine, and
R\"{o}sch \cite{Hu/Staufer/Birkenheuer/Igoshine/Roesch/2000}, Ishida
\cite{Ishida/1998,Ishida/1999,Ishida/2000,Ishida/2002,Ishida/2003},
Kuang and Lin \cite{Kuang/Lin/1997a,Kuang/Lin/1997b}, Maretis
\cite{Maretis/1979},
Matsuoka \cite{Matsuoka/1992a,Matsuoka/1992b,Matsuoka/1998a,%
  Matsuoka/1998b,Matsuoka/2003}, and Saunders \cite{Saunders/1983}).

The usefulness of ${\mathcal{Y}}_{\ell}^{m} (\nabla)$ in connection
with Gaussian functions follows at once from the fact that $\exp
(-\alpha r^2)$ is an eigenfunction of the differential operator $(1/r)
\mathrm{d}/ \mathrm{d} r$ with eigenvalue $-2\alpha$. Thus, the
application of ${\mathcal{Y}}_{\ell}^{m} (\nabla)$ to $\exp
(-\alpha r^2)$ via (\ref{YlmNabla_phi}) yields according to Fieck
\cite{Fieck/1980} the following remarkably compact irreducible spherical
tensor of rank $\ell$:
\begin{equation}
  \label{YlmNabla_GTO}
{\mathcal{Y}}_{\ell}^{m} (\nabla) \, \exp (-\alpha r^2) \; = \;
(-2\alpha)^{\ell} \, \exp (-\alpha r^2) \, 
{\mathcal{Y}}_{\ell}^{m} (\bm{r}) \, .
\end{equation}
A detailed discussion of the usefulness of the differential operator
${\mathcal{Y}}_{\ell}^{m} (\nabla)$ in the context of Gaussian
functions and their multicenter integrals would certainly be of
considerable interest, but unfortunately it would clearly be beyond the
scope of this article. In addition, I cannot claim that molecular
integrals of Gaussian functions are my field of expertise. Therefore,
the interested reader is referred to the articles listed above.

Another class of functions, which have close ties to the differential
operator ${\mathcal{Y}}_{\ell}^{m} (\nabla)$, are the s-called $B$
functions which will be discussed in a detailed way in Section
\ref{Sec:RBF}.

\typeout{==> Section: Fourier Transformation}
\section{Fourier Transformation}
\label{Sec:FourierTr}

As discussed in Section \ref{Sec:HobsonDiffTheor}, Hobson's theorem
implies that a spherical tensor of rank $\ell$ can be generated by
applying ${\mathcal{Y}}_{\ell}^{m} (\nabla)$ to a spherically
symmetric function $\varphi (r)$ according to (\ref{YlmNabla_phi}). If
an irreducible spherical tensor $F_{\ell_2}^{m_2}$ of the type of
(\ref{Def_IrrSphericalTensor}) also satisfies (\ref{Def_Phi_l2m2}), then
the simple explicit expression (\ref{YlmNabla_PHIlm}) for the product
${\mathcal{Y}}_{\ell_1}^{m_1} (\nabla) F_{\ell_2}^{m_2} (\bm{r})$
can be derived via (\ref{YlmNabla_phi}).

In principle, (\ref{YlmNabla_PHIlm}) should suffice for our purposes,
since the scalar function $\Phi_{\ell_2} (r)$ in (\ref{Def_Phi_l2m2})
can according to (\ref{YlmNabla_phi}) be obtained from the scalar
function $f_{\ell_2} (r)$ in (\ref{Def_IrrSphericalTensor}) by repeated
integration with respect to $r$. However, repeated integrations are at
least potentially a source of trouble, and it is thus desirable to
express the product ${\mathcal{Y}}_{\ell_1}^{m_1} (\nabla)
F_{\ell_2}^{m_2} (\bm{r})$ according to (\ref{YlmNab2Flm_GenStruc}) in
terms of radial functions $\gamma _{\ell_1 \ell_2}^{\ell} (r)$ that can
be obtained by differentiating the radial function $f_{\ell_2} (r)$ in
(\ref{Def_IrrSphericalTensor}). That this is indeed possible can be
shown most easily with the help of Fourier transformation, which will be
used in its symmetrical form. Thus, a function $f: \mathbb{R}^3 \to
\mathbb{C}$ and its Fourier transform $\bar{f}$ are connected by the
integrals (\ref{Def_FT}) and (\ref{Def_InvFT}).

The practical usefulness of Fourier transformation is obvious, and it
would clearly be beyond the scope of this article to mention all
successful scientific applications. Let me just mention that Fourier
transformation is -- as first shown by Prosser and Blanchard
\cite{Prosser/Blanchard/1962} and by Geller \cite{Geller/1963b} -- one
of the principal methods of handling molecular multicenter integrals.

For example, the convolution of two functions $f, g: \mathbb{R}^3 \to
\mathbb{C}$ can be expressed as an inverse Fourier integral (see for
example \cite[Eq.\ (5.11)]{Weniger/Steinborn/1983c}):
\begin{equation}
  \label{Convol_fg}
\int \, f (\bm{r}-\bm{r}') \, g (\bm{r}') \, \mathrm{d}^3 \bm{r}' 
\; = \;
\int \, \mathrm{e}^{\mathrm{i} \bm{r} \cdot \bm{p}} \, \bar{f} (\bm{p}) 
\, \bar{g} (\bm{p})  \, \mathrm{d}^3 \bm{p} \, ,
\end{equation}
We immediately obtain this result if we replace in (\ref{Convol_fg}) $f$
and $g$ by their inverse Fourier integrals according to
(\ref{Def_InvFT}) and use the following integral representation of the
three-dimensional delta function (see for example \cite[Eq.\ 
(3.16)]{Judd/1975}):
\begin{equation}
\delta (\bm{r}) \; = \; (2\pi)^{-3} \, \int \,
\mathrm{e}^{\mathrm{i} \bm{r} \cdot \bm{p}} \, \mathrm{d}^3 \bm{p} \, . 
\end{equation}

Six-dimensional integrals describing the Coulomb interaction of two
charge densities $[f (\bm{r}_1)]^{*}$ and $g (\bm{r}_2)$ can also be
expressed as three-dimensional inverse Fourier integrals (see for
example \cite[Eq.\ (2.22)]{Weniger/Grotendorst/Steinborn/1986b}):
\begin{align}
  \label{Coulomb_fg}
  & \int \int \, [f (\bm{r}_1)]^{*} \, \frac{1}{\vert \bm{r}_1 -
    \bm{r}_2 -\bm{R} \vert} \, g (\bm{r}_2) \, \mathrm{d}^3 \bm{r}_1 \,
  \mathrm{d}^3 \bm{r}_2
  \notag \\[1.5\jot]
  & \qquad \; = \; 4\pi \, \int \, \frac{\mathrm{e}^{-\mathrm{i} \bm{R}
      \cdot \bm{p}}}{p^2} \, [\bar{f} (\bm{p})]^{*} \, \bar{g} (\bm{p})
  \, \mathrm{d}^3 \bm{p} \, ,
\end{align}
For a proof of (\ref{Coulomb_fg}), we also need the Fourier transform
(\ref{FT_CoulombPot}) of the Coulomb potential, which -- as discussed in
more details in Section \ref{Sec:SpherDeltaFun} -- holds only in the
sense of distributions.

In connection with multicenter integrals in general and with the
spherical tensor gradient operator in special, Fourier transformation
suffers from a serious limitation which must not be ignored.
Classically, Fourier transformation is defined only for functions that
are absolutely integrable, i.e., which belong to the function space $L^1
(\mathbb{R}^3)$, but by means of a suitable limiting procedure it can be
extended uniquely to give a unitary map from the Hilbert space $L^2
(\mathbb{R}^3)$ of square integrable functions onto itself \cite[Theorem
IX.6 on p.\ 10]{Reed/Simon/1975}. Unfortunately, not all functions of
interest are square integrable and thus possess Fourier transforms that
are meaningful in the sense of classical analysis. As discussed in more
details in Section \ref{Sec:SpherDeltaFun}, the Coulomb potential and
the irregular solid harmonic, which are connected via
(\ref{YllNabla_Coul}), possess Fourier transforms only in the sense of
generalized functions or distributions. For the sake of conceptual
simplicity, let us tacitly assume that all Fourier integrals in this
Section are meaningful in the sense of classical analysis.

Fourier transformation greatly simplifies the treatment of the spherical
tensor gradient operator because of the obvious relationship
\begin{equation}
  \label{YlmNabla_PlaneWave}
{\mathcal{Y}}_{\ell}^{m} (\nabla) \, 
\mathrm{e}^{\mathrm{i} \bm{r} \cdot \bm{p}} \; = \;
\mathrm{i}^{\ell} \, {\mathcal{Y}}_{\ell}^{m} (\bm{p}) \, 
\mathrm{e}^{\mathrm{i} \bm{r} \cdot \bm{p}} \, .
\end{equation}
Thus, in momentum space ${\mathcal{Y}}_{\ell}^{m} (\nabla)$ is a simple
multiplicative operator with known coupling properties. Moreover,
(\ref{YlmNabla_PlaneWave}) makes it plausible that
${\mathcal{Y}}_{\ell}^{m} (\nabla)$ is indeed an irreducible spherical
tensor of rank $\ell$ (compare \cite[p.\ 312]{Biedenharn/Louck/1981a}).

If we use in the Fourier integral (\ref{Def_FT}) both
(\ref{Def_IrrSphericalTensor}) as well as the Rayleigh expansion of a
plane wave in terms of spherical Bessel functions and spherical
harmonics (compare for instance \cite[p.\ 442]{Biedenharn/Louck/1981a}),
\begin{equation}
  \label{Rayleigh_Expan}
\mathrm{e}^{\pm \mathrm{i} \bm{p} \cdot \bm{r}} \; = \; 
4\pi \, \sum_{\ell=0}^{\infty} \, (\pm 1)^{\ell} \, j_{\ell} (pr) \,
\sum_{m=-\ell}^{\ell} \, \bigl[ Y_{\ell}^{m} (\bm{p}/p) \bigr]^{*} \,
Y_{\ell}^{m} (\bm{r}/r) \, ,
\end{equation}
we see that the Fourier transform $\bar{F}_{\ell_2}^{m_2} (\bm{p})$ of
the irreducible spherical tensor $F_{\ell_2}^{m_2} (\bm{r})$ is also a
spherical tensor of rank $\ell_2$ in momentum space since it can be
expressed as a spherical harmonic multiplied by a radial integral
\cite[Eqs.\ (4.3) and (4.4)]{Weniger/Steinborn/1983c}:
\begin{subequations}
   \label{FT_Flm}
\begin{align}
  \label{FT_Flm_a}
  \bar{F}_{\ell_2}^{m_2} (\bm{p}) & \; = \; \bar{f}_{\ell_2} (p) \,
  Y_{\ell_2}^{m_2} (\bm{p}/p) \, ,
  \\
  \label{FT_Flm_b}
  \bar{f}_{\ell_2} (p) & \; = \; (-\mathrm{i})^{\ell_2} \, p^{-1/2} \,
  \int_{0}^{\infty} \, r^{3/2} \, J_{\ell_2+1/2} (pr) \, f_{\ell_2} (r)
  \, \mathrm{d} r \,.
\end{align}  
\end{subequations}
In the same way, we obtain the following Hankel-type integral
representation for the radial part $f_{\ell_2} (r)$ of $F_{\ell_2}^{m_2}
(\bm{r})$ \cite[Eq.\ (4.5)]{Weniger/Steinborn/1983c}:
\begin{equation}
  \label{IntRep_f_l(r)}
f_{\ell_2} (r) \; = \; \mathrm{i}^{\ell_2} \, r^{-1/2} \,
\int_{0}^{\infty} \, p^{3/2} \, J_{\ell_2+1/2} (rp) \, 
\bar{f}_{\ell_2} (p) \, \mathrm{d} p \, .
\end{equation}

Let us now consider the product ${\mathcal{Y}}_{\ell_1}^{m_1}
(\nabla) F_{\ell_2}^{m_2} (\bm{r})$. With the help of
(\ref{YlmNabla_PlaneWave}) and (\ref{Def_InvFT}), we obtain the
following Fourier integral representation:
\begin{equation}
{\mathcal{Y}}_{\ell_1}^{m_1} (\nabla) \, F_{\ell_2}^{m_2} (\bm{r}) 
\; = \; \mathrm{i}^{\ell_1} \, (2\pi)^{-3/2} \, 
\int \, \mathrm{e}^{\mathrm{i} \bm{r}\cdot\bm{p}} 
\, {\mathcal{Y}}_{\ell_1}^{m_1} (\bm{p}) \,
\bar{F}_{\ell_2}^{m_2} (\bm{p}) \, \mathrm{d}^3 \bm{p} \, .
\end{equation}
Next, we use the Rayleigh expansion (\ref{Rayleigh_Expan}) and replace
the spherical harmonics by Gaunt coefficients according to
(\ref{Ylm_lin}). This yields \cite[Eq.\ (4.7)]{Weniger/Steinborn/1983c}:
\begin{align}
  \label{FTIntRep_YlnNabla2Flm}
  & {\mathcal{Y}}_{\ell_1}^{m_1} (\nabla) \, F_{\ell_2}^{m_2}
  (\bm{r})
  \notag \\
  & \qquad \; = \;
  \sum_{\ell=\ell_{\mathrm{min}}}^{\ell_{\mathrm{max}}} \! {}^{(2)}
  \, \langle \ell m_1+m_2 \vert \ell_1 m_1 \vert \ell_2 m_2 \rangle \,
  \mathrm{i}^{\ell+\ell_1} \, Y_{\ell}^{m_1+m_2} (\bm{r}/r)
  \notag \\
  & \qquad \qquad \times \, r^{-1/2} \, \int_{0}^{\infty} \,
  p^{\ell_1+3/2} \, J_{\ell+1/2} (rp) \, \bar{f}_{\ell_2} (p) \,
  \mathrm{d} p \, .
\end{align}
Comparison of (\ref{YlmNab2Flm_GenStruc}) and
(\ref{FTIntRep_YlnNabla2Flm}) yields the following integral
representation for the radial function $\gamma _{\ell_1 \ell_2}^{\ell}
(r)$ \cite[Eq.\ (4.8)]{Weniger/Steinborn/1983c}:
\begin{equation}
  \label{IntRep_gamma}
\gamma _{\ell_1 \ell_2}^{\ell} (r) \; = \; \mathrm{i}^{\ell+\ell_1} \,
r^{-1/2} \, \int_{0}^{\infty} \, p^{\ell_1+3/2} \, J_{\ell+1/2} (rp) \, 
\bar{f}_{\ell_2} (p) \, \mathrm{d} p \, .
\end{equation}
Now, all we need is a differential operator in $r$ that generates the
integral in (\ref{IntRep_gamma}) from the integral in
(\ref{IntRep_f_l(r)}). With the help of known properties of the Bessel
function $J_{\nu} (z)$ or by direct differentiation techniques (see
\cite[Section III]{Weniger/Steinborn/1983c}), this can be accomplished
relatively easily \cite[Eqs.\ (3.29), (4.15) - (4.18), and
(4.24)]{Weniger/Steinborn/1983c}: {\allowdisplaybreaks
\begin{align}
  \label{f2gamma_1}
  \gamma _{\ell_1 \ell_2}^{\ell} (r) & \; = \; \sum_{q=0}^{\Delta \ell}
  \, \frac{(-\Delta \ell)_q (-\sigma (\ell)-1/2)_q} {q!} \, 2^q \,
  r^{\ell_1 + \ell_2 - 2 q}
  \notag \\
  & \qquad \times \, \left( \frac{1}{r} \frac{\mathrm{d}}{\mathrm{d} r}
  \right)^{\ell_1 - q} \, \frac{f_{\ell_2} (r)}{r^{\ell_2}}
  \\[1.5\jot]
  \label{f2gamma_2}
  & \; = \; r^{-\ell-1} \, \left( \frac{1}{r}
    \frac{\mathrm{d}}{\mathrm{d} r} \right)^{\Delta \ell} \,
  r^{\ell_1+\ell_2+\ell+1} \, \left( \frac{1}{r}
    \frac{\mathrm{d}}{\mathrm{d} r} \right)^{\Delta \ell_2} \,
  \frac{f_{\ell_2} (r)}{r^{\ell_2}}
  \\[1.8\jot]
  \label{f2gamma_3}
  & \; = \; r^{\ell} \, \left( \frac{1}{r} \frac{\mathrm{d}}{\mathrm{d}
      r} \right)^{\Delta \ell_2} \, r^{\ell_1-\ell_2-\ell-1} \, \left(
    \frac{1}{r} \frac{\mathrm{d}}{\mathrm{d} r} \right)^{\Delta \ell} \,
  r^{\ell_2+1} \, f_{\ell_2} (r)
  \\[1.9\jot]
  \label{f2gamma_4}
  & \; = \; r^{-\ell-1} \, \left( \frac{1}{r}
    \frac{\mathrm{d}}{\mathrm{d} r} \right)^{\Delta \ell_2} \,
  r^{\ell_1-\ell_2+3\ell+1} \, \left( \frac{1}{r}
    \frac{\mathrm{d}}{\mathrm{d} r} \right)^{\Delta
    \ell_2} \notag \\[1.5\jot]
  & \qquad \times \, r^{-2\ell-1} \, \left( \frac{1}{r}
    \frac{\mathrm{d}}{\mathrm{d} r} \right)^{\ell_2-\ell} r^{\ell_2+1}
  \, f_{\ell_2} (r)
  \\[1.5\jot]
  \label{f2gamma_5}
  & \; = \; r^{\ell} \, \left( \frac{1}{r} \frac{\mathrm{d}}{\mathrm{d}
      r} \right)^{\Delta \ell} \, r^{\ell_1+\ell_2-3\ell-1} \, \left(
    \frac{1}{r} \frac{\mathrm{d}}{\mathrm{d} r} \right)^{\Delta
    \ell} \notag \\[1.5\jot]
  & \qquad \times \, r^{2\ell+1} \, \left( \frac{1}{r}
    \frac{\mathrm{d}}{\mathrm{d} r} \right)^{\ell-\ell_2}
  \frac{f_{\ell_2} (r)}{r^{\ell_2}}
  \\[1.5\jot]
  \label{f2gamma_6}
  & \; = \; \sum_{s=0}^{\Delta \ell_2} \, \frac{(-\Delta \ell_2)_s
    (\Delta \ell_1 + 1/2)_s}{s!} \,
  2^s \, r^{\ell_1 - \ell_2 - 2 s - 1}\notag \\[1.5\jot]
  & \qquad \times \, \left( \frac{1}{r} \frac{\mathrm{d}}{\mathrm{d} r}
  \right)^{\ell_1 - s} \, r^{\ell_2 + 1} \, f_{\ell_2} (r) \, .
\end{align}
}%
The abbreviations $\Delta l$, $\Delta l_1$, $\Delta l_2$, and $\sigma
(\ell)$ are defined in (\ref{Def_Del_l}) -- (\ref{Def_sigma_l}).
Obviously, the representations (\ref{f2gamma_4}) and (\ref{f2gamma_5})
make sense only if either $\ell_2 \ge \ell$ or $\ell \ge \ell_2$ hold.

The expressions (\ref{f2gamma_1}) -- (\ref{f2gamma_6}) for $\gamma
_{\ell_1 \ell_2}^{\ell} (r)$ all look quite different. Nevertheless,
their equivalence can be shown explicitly with the help of summation
theorems for generalized hypergeometric series with unit argument (see
the discussion in connection with \cite[Eqs.\ (4.19) --
(4.24)]{Weniger/Steinborn/1983c}).

Alternative expressions for $\gamma _{\ell_1 \ell_2}^{\ell} (r)$ as well
as for more general radial functions, that occur in the case of the
product $\nabla^{2n} {\mathcal{Y}}_{\ell_1}^{m_2} (\nabla)
F_{\ell_2}^{m_2} (\bm{r})$ with $n \in \mathbb{N}_0$, were considered by
Santos \cite{Santos/1973}, Bayman \cite{Bayman/1978}, Stuart
\cite{Stuart/1981}, Niukkanen \cite{Niukkanen/1983a}, Weniger and
Steinborn \cite{Weniger/Steinborn/1983c}, and Rashid \cite{Rashid/1986}.

The expressions (\ref{f2gamma_1}) -- (\ref{f2gamma_6}) for $\gamma
_{\ell_1 \ell_2}^{\ell} (r)$ all have a manageable complexity and are
well suited for practical applications. Nevertheless, it is more
convenient to compute the product ${\mathcal{Y}}_{\ell_1}^{m_2}
(\nabla) F_{\ell_2}^{m_2} (\bm{r})$ or also the more general
product $\nabla^{2n} {\mathcal{Y}}_{\ell_1}^{m_2} (\nabla)
F_{\ell_2}^{m_2} (\bm{r})$ with $n \in \mathbb{N}_0$ via
(\ref{YlmNabla_PHIlm}) if the function $\Phi_{\ell_2} (r)$ defined in
(\ref{Def_Phi_l2m2}) is easily accessible and has a sufficiently simple
form as in the case of the irregular solid harmonic and the Gaussian
function according to (\ref{YllNabla_Coul}) and (\ref{YlmNabla_GTO}),
respectively, or in the case of $B$ functions according to
(\ref{STGO_Bn00}). 

With the help of Fourier transformation, it is easy to obtain the
explicit expression for the product $\mathcal{Y}_{\ell}^{m}
(\nabla) f (\bm{r}) g (\bm{r})$ which could be called the Leibniz
theorem of the spherical tensor gradient operator and which was
originally derived by Dunlap \cite[Eq.\ (10)]{Dunlap/1990}, albeit in a
somewhat cryptic way. My derivation is inspired by Grotendorst and
Steinborn \cite[Appendix A]{Grotendorst/Steinborn/1985} who had derived
a more general expression involving spherical tensor gradient operators
in connection with the Fourier transform of a two-center density.

For that purpose, let us express both $f$ and $g$ as inverse Fourier
integrals according to (\ref{Def_InvFT}) and use
(\ref{YlmNabla_PlaneWave}):
\begin{align}
  & \mathcal{Y}_{\ell}^{m} (\nabla) \, f (\bm{r}) \, g (\bm{r}) \;
  = \; \frac{\mathcal{Y}_{\ell}^{m} (\nabla)}{(2\pi)^{3}} \, \int
  \int \, \mathrm{e}^{\mathrm{i} \bm{r} \cdot (\bm{p}+\bm{q})} \,
  \bar{f} (\bm{p}) \, \bar{g} (\bm{p}) \, \mathrm{d}^3 \bm{p} \,
  \mathrm{d}^3 \bm{q}
  \\
  \label{IntRep_YlmNabla_fg}
  & \qquad \; = \; (2\pi)^{-3} \, \int \int \, \mathrm{e}^{\mathrm{i}
    \bm{r} \cdot (\bm{p}+\bm{q})} \, \mathcal{Y}_{\ell}^{m} (
  \mathrm{i}[\bm{p}+\bm{q}]) \, \bar{f} (\bm{p}) \, \bar{g} (\bm{p}) \,
  \mathrm{d}^3 \bm{p} \, \mathrm{d}^3 \bm{q} \, .
\end{align}
The dependence of the regular spherical harmonic on the two momentum
variables $\bm{p}$ and $\bm{q}$ can be decoupled with the help of the
well known addition theorem of the regular solid harmonics (a
particularly simple derivation of this addition theorem based on the the
spherical tensor gradient operator can be found in 
\cite[Section 6]{Weniger/2000a}):
\begin{align}
  \label{AddTh_Ylm}
  \mathcal{Y}_{\ell}^{m} (\bm{r} + \bm{r}') & \; = \;
  \sum_{\lambda=0}^{\ell} \, \frac{2\pi \,
    (1/2)_{\ell+1}}{(1/2)_{\lambda+1} (1/2)_{\ell-\lambda+1}}
  \notag \\
  & \qquad \times \, \sum_{\mu = - \lambda}^{\lambda} \, \langle \ell \,
  m \vert \lambda \, - \mu \vert \ell-\lambda \, m+\mu \rangle \,
  \mathcal{Y}_{\lambda}^{-\mu} (\bm{r}) \,
  \mathcal{Y}_{\ell-\lambda}^{m+\mu} (\bm{r}') \, .
\end{align}
The Gaunt coefficients in this addition theorem can be expressed in
closed form (compare for instance \cite[Eq.\ (6.5)]{Weniger/2000a} which
unfortunately contains a typographical error: The factor
$(2\ell_2-\ell_1+1)$ in the numerator of the square root on the
right-hand side has to be replaced by $(2\ell_2-2\ell_1+1)$.) In this
way, we obtain Steinborn's factor-free version of the addition theorem
of the regular solid harmonics \cite[Eq.\ (9)]{Steinborn/1969}.

If we now insert (\ref{AddTh_Ylm}) into (\ref{IntRep_YlmNabla_fg}) and
use (\ref{YlmNabla_PlaneWave}), we obtain: {\allowdisplaybreaks
\begin{align}
  & \mathcal{Y}_{\ell}^{m} (\nabla) \, f (\bm{r}) \, g (\bm{r})
  \notag \\
  & \qquad \; = \; \sum_{\lambda=0}^{\ell} \, \frac{2\pi \,
    (1/2)_{\ell+1}}{(1/2)_{\lambda+1} (1/2)_{\ell-\lambda+1}} \,
  \sum_{\mu = - \lambda}^{\lambda} \, \langle \ell \, m \vert \lambda \,
  - \mu \vert \ell-\lambda \, m+\mu \rangle
  \notag \\
  & \qquad \qquad \times (2\pi)^{-3/2} \, \int \, \mathrm{e}^{\mathrm{i}
    \bm{r} \cdot \bm{p}} \, \mathcal{Y}_{\lambda}^{-\mu} (
  \mathrm{i}\bm{p}) \, \bar{f} (\bm{p}) \, \mathrm{d}^3 \bm{p}
  \notag \\
  & \qquad \qquad \times (2\pi)^{-3/2} \, \int \, \mathrm{e}^{\mathrm{i}
    \bm{r} \cdot \bm{q}} \, \mathcal{Y}_{\ell-\lambda}^{m+\mu} (
  \mathrm{i}\bm{q}) \, \bar{g} (\bm{q}) \, \mathrm{d}^3 \bm{q} \\
  & \qquad \; = \; \sum_{\lambda=0}^{\ell} \, \frac{2\pi \,
    (1/2)_{\ell+1}}{(1/2)_{\lambda+1} (1/2)_{\ell-\lambda+1}} \,
  \sum_{\mu = - \lambda}^{\lambda} \, \langle \ell \, m \vert \lambda \,
  - \mu \vert \ell-\lambda \, m+\mu \rangle
  \notag \\
  & \qquad \qquad \times \biggl[ (2\pi)^{-3/2} \,
  \mathcal{Y}_{\lambda}^{-\mu} (\nabla) \, \int \,
  \mathrm{e}^{\mathrm{i} \bm{r} \cdot \bm{p}} \, \bar{f} (\bm{p}) \,
  \mathrm{d}^3 \bm{p} \biggr]
  \notag \\[1.5\jot]
  & \qquad \qquad \times \biggl[ (2\pi)^{-3/2} \,
  \mathcal{Y}_{\ell-\lambda}^{m+\mu} (\nabla) \, \int \,
  \mathrm{e}^{\mathrm{i} \bm{r} \cdot \bm{q}} \,
  \bar{g} (\bm{q}) \, \mathrm{d}^3 \bm{q} \biggr] \\
  \label{YlmNabla_fg}
  & \qquad \; = \; \sum_{\lambda=0}^{\ell} \, \frac{2\pi \,
    (1/2)_{\ell+1}}{(1/2)_{\lambda+1} (1/2)_{\ell-\lambda+1}} \,
  \sum_{\mu = - \lambda}^{\lambda} \, \langle \ell \, m \vert \lambda \,
  - \mu \vert \ell-\lambda \, m+\mu \rangle
  \notag \\
  & \qquad \qquad \times \bigl[ \mathcal{Y}_{\lambda}^{-\mu}
  (\nabla) \, f (\bm{r}) \bigr] \, \bigl[
  \mathcal{Y}_{\ell-\lambda}^{m+\mu} (\nabla) \, g (\bm{r}) \bigr]
  \, .
\end{align}
} Thus, the Leibniz theorem (\ref{YlmNabla_fg}) of the spherical tensor
gradient operator is nothing but the addition theorem (\ref{AddTh_Ylm})
of the regular solid harmonic in momentum space.

\typeout{==> Section: Reduced Bessel Functions} 
\section{Reduced Bessel Functions}
\label{Sec:RBF}

In Section \ref{Sec:HobsonDiffTheor} it was shown that the application
of the spherical tensor gradient operator ${\mathcal{Y}}_{\ell}^{m}
(\nabla)$ to the Coulomb potential or to a scalar Gaussian yields
remarkably compact results according to (\ref{YllNabla_Coul}) or
(\ref{YlmNabla_GTO}), respectively.

Other functions, to which the spherical tensor gradient operator can be
applied with remarkable ease, are the so-called reduced Bessel
functions, whose use in quantum chemistry had been proposed by Shavitt
\cite[Eq.\ (55) on p.\ 15]{Shavitt/1963}, and their anisotropic
generalizations, the so-called $B$ functions, which had been introduced
by Filter and Steinborn \cite[Eq.\ (2.14)]{Filter/Steinborn/1978b}.
Detailed discussions of the mathematical properties of reduced Bessel
functions and $B$ functions can be found in my PhD thesis
\cite{Weniger/1982} and in the PhD thesis of Homeier
\cite{Homeier/1990}.

If $K_{\nu} (z)$ is a modified Bessel function of the second kind
\cite[p.\ 66]{Magnus/Oberhettinger/Soni/1966}, the reduced Bessel
function is defined by \cite[Eqs.\ (3.1) and
(3.2)]{Steinborn/Filter/1975c}
\begin{equation}
   \label{Def:RBF}
\hat{k}_{\nu} (z) \; = \; (2/\pi)^{1/2} \, z^{\nu} \, K_{\nu} (z) \, .
\end{equation}
If the order $\nu$ of a reduced Bessel function is half-integral, $\nu =
n + 1/2$ with $n \in \mathbb{N}_0$, the reduced Bessel function can be
written as an exponential multiplied by a terminating confluent
hypergeometric series ${}_1 F_1$ \cite[Eq.\ 
(3.7)]{Weniger/Steinborn/1983b}:
\begin{equation}
  \label{RBF_HalfInt}
\hat{k}_{n+1/2} (z) \; = \; 
2^n \, (1/2)_n \, \mathrm{e}^{-z} \, {}_1 F_1 (-n; -2n; 2z) \, .
\end{equation}
The polynomial part in (\ref{RBF_HalfInt}) was also treated
independently in the mathematical literature, where the notation
$\Theta_n (z) = \mathrm{e}^z \, \hat{k}_{n+1/2} (z)$ is used. Together
with some other, closely related polynomials, the $\Theta_n (z)$ are
called Bessel polynomials \cite{Grosswald/1978}. According to Grosswald
\cite{Grosswald/1978}, they are applied in such diverse fields as number
theory, statistics, and the analysis of complex electrical networks.

The Bessel polynomials $\Theta_n (z)$ occur also in a completely
different mathematical context: In the book by Baker and Graves-Morris
\cite[p.\ 8]{Baker/Graves-Morris/1996} on Pad\'{e} approximants, it is
remarked that Pad\'{e} had shown in his seminal thesis \cite{Pade/1892}
that the Pad\'{e} approximant $[n/m]$ to the exponential function $\exp
(z)$ can be expressed as the ratio of two terminating confluent
hypergeometric series \cite[Eq.\ (2.12)]{Baker/Graves-Morris/1996}:
\begin{equation}
[n/m] \; = \; 
\frac{{}_1 F_1 (- n; - n - m; z)}{{}_1 F_1 (- n; - n - m; - z)} \, ,
\qquad n, m \in \mathbb{N}_0 \, .
\end{equation}
Accordingly, the diagonal Pad\'{e} approximant with $n = m$ to the
exponential function can be expressed as the ratio of two Bessel
polynomials:
\begin{equation}
[n/n] \; = \; \frac{\theta_n (z/2)}{\theta_n (-z/2)} \, ,
\qquad n \in \mathbb{B}_0 \, .
\end{equation}

As an anisotropic generalization of the reduced Bessel function with
half-integral order, the so-called $B$ function was introduced by Filter
and Steinborn \cite[Eq.\ (2.14)]{Filter/Steinborn/1978b},
\begin{equation}
  \label{Def:B_Fun}
B_{n,\ell}^{m} (\alpha, \hm{r}) \; = \; 
[2^{n+\ell} (n+\ell)!]^{-1} \, \hat{k}_{n-1/2} (\alpha r) \,
\mathcal{Y}_{\ell}^{m} (\alpha \bm{r}) \, , 
\quad \alpha > 0 \, , n \in \mathbb{Z} \, .
\end{equation}

$B$ functions are a fairly large class of exponentially decaying
functions. For $n \in \mathbb{N}$, they are suited to serve as trial
functions in LCAO-MO calculations. Since, however, $B$ functions have a
much more complicated mathematical structure than for example
Slater-type functions, whose molecular multicenter integrals functions
are notoriously difficult, it is by no means obvious that anything can
be gained by using $B$ functions instead of Slater-type functions as
basis functions. However, $B$ functions possess a Fourier transform of
remarkable simplicity:
\begin{align}
  \label{FT_B_Fun}
  \bar{B}_{n,\ell}^{m} (\alpha, \bm{p}) & \; = \; (2\pi)^{-3/2} \, \int 
  \, \mathrm{e}^{- \mathrm{i} \bm{p} \cdot \bm{r}} \, B_{n,\ell}^{m}
  (\alpha, \bm{r}) \, \mathrm{d}^3 \bm{r}
  \notag \\
  & \; = \; (2/\pi)^{1/2} \, \frac{\alpha^{2n+\ell-1}}{[\alpha^2 +
    p^2]^{n+\ell+1}} \, \mathcal{Y}_{\ell}^{m} (- i \bm{p}) \, .
\end{align}
This is most likely the most consequential and certainly the most often
cited result of my PhD thesis \cite[Eq.\ (7.1-6) on p.\ 
160]{Weniger/1982}. Later, (\ref{FT_B_Fun}) was published in \cite[Eq.\ 
(3.7)]{Weniger/Steinborn/1983a}. Independently and almost
simultaneously, the Fourier transform (\ref{FT_B_Fun}) was also derived
by Niukkanen \cite{Niukkanen/1984c}.

The exceptionally simple Fourier transform (\ref{FT_B_Fun}) gives $B$
functions a unique position among exponentially decaying basis
functions, and it also explains why other exponentially decaying
functions like Slater-type functions, bound state hydrogen
eigenfunctions or other functions sets based on generalized Laguerre
polynomials can all be expressed in terms of finite linear combinations
of $B$ functions (details and further references can found in
\cite[Section IV]{Weniger/1985} or \cite[Section 4]{Weniger/2002}).

As discussed in Section \ref{Sec:FourierTr}, Fourier transformation is
one of the principal approaches for the treatment of multicenter
integrals \cite{Prosser/Blanchard/1962,Geller/1963b}. 

Thus, the simplicity of the Fourier transform (\ref{FT_B_Fun}) makes it
plausible that multicenter integrals of $B$ functions can normally be
evaluated more easily than the analogous integrals of other
exponentially decaying functions as for instance Slater-type functions,
For example, by inserting (\ref{FT_B_Fun}) into the inverse Fourier
integral on the right-hand side of (\ref{Convol_fg}), we immediately
obtain the following, extremely simple expression for the convolution
integral of two $B$ functions with equal scaling parameters:
{\allowdisplaybreaks
\begin{align}
  \label{Conv_Bnlm_alpha}
& \int \, B_{n_1,\ell_1}^{m_1} (\alpha, [\bm{r}-\bm{r}']) \,
  B_{n_2, \ell_2}^{m_2} (\alpha, \bm{r}') \, \mathrm{d} \bm{r}'
\notag \\ 
& \qquad \; = \; \frac{4\pi}{\alpha^3} \,
  \sum_{\ell=\ell_{\mathrm{min}}}^{\ell_{\mathrm{max}}} \! {}^{(2)}
  \, \langle \ell m_1+m_2 \vert \ell_1 m_1 \vert \ell_2 m_2 \rangle
  \notag \\
  & \qquad \qquad \times \, \sum_{t=0}^{\Delta \ell} \, (-1)^t \,
  {\binom{\Delta \ell} {t}} \, B_{n_1+n_2+\ell_1+\ell_2-\ell-t+1, 
  \ell}^{m_1+m_2} (\alpha, \bm{r}) \, .
\end{align}
}%
This expression was originally derived by Filter and Steinborn
\cite[Eq.\ (4.1)]{Filter/Steinborn/1978a} with the help of an addition
theorem. The summation limits $\ell_{\mathrm{min}}$ and
$\ell_{\mathrm{max}}$ are given in (\ref{SumLims}), and $\Delta \ell$ is
defined by (\ref{Def_Del_l}).

In recent years, some significant progress has been achieved with
respect to molecular multicenter integrals of $B$ functions (see for
example the articles by Steinborn, Ho\-meier, Fern\'andez Rico, Ema,
L\'opez, and Ram\'{\i}rez
\cite{Steinborn/Homeier/FernandezRico/Ema/Lopez/Ramirez/1999}, and
Steinborn, Ho\-meier, Ema, L\'opez, and Ram\'{\i}rez
\cite{Steinborn/Homeier/Ema/Lopez/Ramirez/2000}). Particularly promising
seems to be the approach of Safouhi who -- starting from the Fourier
transform (\ref{FT_B_Fun}) -- converts complicated multicenter integrals
of $B$ or Slater-type functions to multi-dimensional integral
representations involving nonphysical variables that have to be
evaluated by numerical quadrature. At first sight, this does not look
like a good idea because of the oscillatory nature of the
multi-dimensional integral representations, which makes the
straightforward application of conventional quadrature methods
difficult. However, these computational problems can be overcome if the
quadrature schemes are combined with suitable nonlinear extrapolation
methods. Based on previous work of Sidi\cite{Sidi/1980c} and of Levin
and Sidi \cite{Levin/Sidi/1981} on extrapolation methods for numerical
quadrature schemes, Safouhi succeeded in developing some extrapolation
techniques specially suited for his needs. This permits a remarkably
efficient and reliable evaluation of complicated molecular multicenter
integrals via oscillatory (Fourier based) integral representations (see
for example the recent articles by Berlu and Safouhi
\cite{Berlu/Safouhi/2003a,Berlu/Safouhi/2003b,Berlu/Safouhi/2004a},
Berlu, Safouhi and Hoggan \cite{Berlu/Safouhi/Hoggan/2004}, Safouhi
\cite{Safouhi/2000,Safouhi/2001a,Safouhi/2001b,Safouhi/2001c,%
  Safouhi/2002a,Safouhi/2002b}, Safouhi and Hoggan
\cite{Safouhi/Hoggan/2001,Safouhi/Hoggan/2002,Safouhi/Hoggan/2003} and
references therein). Safouhi's work can also be viewed as a convincing
demonstration of the practical usefulness of extrapolation and
convergence acceleration techniques in molecular electronic structure
theory.

It follows at once from (\ref{FT_B_Fun}) that a $B$ function can be
expressed as an inverse Fourier integral according to
\begin{equation}
  \label{InvFT_B_Fun}
B_{n,\ell}^{m} (\alpha, \bm{r}) \; = \;
\frac{\alpha^{2n+\ell-1}}{2\pi^2} 
\, \int \, \mathrm{e}^{\mathrm{i} \bm{r} \cdot \bm{p}} \, \frac
{\mathcal{Y}_{\ell}^{m} (-\mathrm{i} \bm{p})}
{[\alpha^2 + p^2]^{n+\ell+1}} \, \mathrm{d}^3 \bm{p} \, .
\end{equation}
Comparison of (\ref{YlmNabla_PlaneWave}) and (\ref{InvFT_B_Fun}) shows
that the application of the spherical tensor gradient operator to a
scalar $B$ function yields a nonscalar $B$ function \cite[Eq.\ (7.1-10)
on p.\ 161]{Weniger/1982}:
\begin{equation}
   \label{STGO_Bn00}
B_{n,\ell}^{m} (\alpha, \bm{r}) \; = \; 
(4\pi)^{1/2} \, (-\alpha)^{-\ell} \, \mathcal{Y}_{\ell}^{m} (\nabla) \, 
B_{n+\ell,0}^{0} (\alpha, \bm{r}) \, .
\end{equation}
This as well as several other related results can also be deduced from
known properties of the modified Bessel function $K_{\nu} (z)$ via
(\ref{YlmNabla_phi}) (see for example \cite[Eq.\ 
(4.12)]{Weniger/Steinborn/1983a}):

It is also quite easy to derive an explicit expression for the product
$\mathcal{Y}_{\ell_1}^{m_1} (\nabla) B_{n_2,\ell_2}^{m_2} (\bm{r})$ via
(\ref{YlmNabla_PHIlm}) since the application of higher powers of the
Laplacian $\nabla^2$ to a $B$ function poses no problems. It follows at
once from the integral representation (\ref{InvFT_B_Fun}) that the
differential operator $1 - \alpha^{-2} \nabla^2$ of the modified
Helmholtz equation functions as a ladder operator \cite[Eq.\ 
(5.6)]{Weniger/Steinborn/1983c}:
\begin{equation}
  \label{Shift_n_Blm}
[1 - \alpha^{-2} \nabla^2] \, B_{n,\ell}^{m} (\alpha, \bm{r}) \; = \; 
B_{n-1,\ell}^{m} (\alpha, \bm{r}) \, .
\end{equation}
As discussed in more details in Section \ref{Sec:SpherDeltaFun}, this
relationship holds also if the indices $n$ and $\ell$ satisfies $n+\ell
< 0$, i.e., for $B$ functions that are derivatives of the
three-dimensional delta function according to (\ref{Def_Distrib_B_Fun}).

Thus, the binomial expansion of $\alpha^{-2\nu} \, \nabla^{2\nu}$ in
powers of $1 - \alpha^{-2} \nabla^2$ in combination with
(\ref{Shift_n_Blm}) yields \cite[Eq.\ (5.7)]{Weniger/Steinborn/1983c}:
\begin{equation}
  \label{Delta2nB_Fun}
\alpha^{-2\nu} \, \nabla^{2\nu} \, B_{n,\ell}^{m} (\alpha, \bm{r}) 
\; = \; \sum_{t=0}^{\nu} \, (-1)^t \, {\binom{\nu} {t}} \,
B_{n-t,\ell}^{m} (\alpha, \bm{r}) \, .
\end{equation}
If we now combine (\ref{YlmNabla_PHIlm}) with (\ref{Delta2nB_Fun}), 
we immediately obtain the following compact linear combination of Gaunt
coefficients, binomial coefficients, and $B$ functions \cite[Eq.\ 
(6.25)]{Weniger/Steinborn/1983c}:
\begin{align}
  \label{STGO_Bnlm}
  \mathcal{Y}_{\ell_1}^{m_1} (\nabla) \, B_{n_2,\ell_2}^{m_2}
  (\alpha, \bm{r}) & \; = \; (-\alpha)^{\ell_1} \,
  \sum_{\ell=\ell_{\mathrm{min}}}^{\ell_{\mathrm{max}}} \! {}^{(2)}
  \, \langle \ell m_1+m_2 \vert \ell_1 m_1 \vert \ell_2 m_2 \rangle
  \notag \\
  & \qquad \times \, \sum_{t=0}^{\Delta \ell} \, (-1)^t \,
  {\binom{\Delta \ell} {t}} \, B_{n_2+\ell_2-\ell-t,\ell}^{m_1+m_2}
  (\alpha, \bm{r}) \, .
\end{align}
It follows from the summation limits (\ref{SumLims}) that $\Delta \ell$
defined in (\ref{Def_Del_l}) is either a positive integer or zero.

As already remarked above, all the commonly used exponentially decaying
functions like Slater-type functions, bound state hydrogen
eigenfunctions or other functions sets based on generalized Laguerre
polynomials can all be expressed in terms of finite linear combinations
of $B$ functions (see for example \cite[Section IV]{Weniger/1985} or
\cite[Section 4]{Weniger/2002}). Thus, it follows from (\ref{STGO_Bnlm})
that the product of a spherical tensor gradient operator and one of
these exponentially decaying functions can be expressed as a finite
linear combination of $B$ functions.

\typeout{==> Section: Spherical Delta Functions}
\section{Spherical Delta Functions}
\label{Sec:SpherDeltaFun}

Classically, the domain of the spherical tensor gradient operator
consists of the differentiable functions $f : \mathbb{R}^3 \to
\mathbb{C}$, although we are in practice only interested in
differentiable irreducible spherical tensors of the type of
(\ref{Def_IrrSphericalTensor}). However, as for example argued in
Dirac's classic book \cite[\S 15]{Dirac/1958}, the functions defined in
the sense of classical analysis do not suffice in quantum theory. It is
necessary to use also more general mathematical objects, the so-called
generalized functions or distributions, whose mathematical theory was
rigorously formulated by Schwartz (see \cite{Schwartz/1966a} and
references therein).

The best known nonclassical generalization of a function $f :
\mathbb{R}^3 \to \mathbb{C}$ is the three-dimensional Dirac delta
function that can be defined as a generalized solution of the Poisson
equation of a unit point charge \cite[Eq.\ (1.31)]{Jackson/1975}:
\begin{equation}
  \label{PoissonEq_CP}
\nabla^2 \, \frac{1}{r} \; = \; - 4\pi \, \delta (\bm{r}) \, .
\end{equation}
This Poisson equation also expresses the well known fact that the
Coulomb potential is the Green's function of the three-dimensional
Laplace equation.

The prototype of a distribution, which is also an irreducible spherical
tensor of rank $\ell$, is the so-called spherical delta function (see
for example \cite[Eq.\ (30)]{Rowe/1978}):
\begin{equation}
  \label{Def_delta_lm}
\delta_{\ell}^{m} (\bm{r}) \; = \; 
\frac {(-1)^{\ell}} {(2\ell - 1)!!} \,
\mathcal{Y}_{\ell}^{m} (\nabla) \, \delta (\bm{r}) \, .
\end{equation}
The spherical delta function can also be obtained by applying the
Laplacian to an irregular solid harmonics (see for example \cite[Eq.\ 
(29)]{Rowe/1978}):
\begin{equation}
  \label{PoissonEq_Zlm}
\nabla^2 \, \mathcal{Z}_{\ell}^{m} (\bm{r}) \; = \;
- 4 \pi \, \delta_{\ell}^{m} (\bm{r}) \, .
\end{equation}
This follows at once from (\ref{YllNabla_Coul}), (\ref{PoissonEq_CP}),
(\ref{Def_delta_lm}), and the fact that $\nabla^2$ and
$\mathcal{Y}_{\ell}^{m} (\nabla)$ commute. Comparison of
(\ref{PoissonEq_CP}) and (\ref{PoissonEq_Zlm}) shows that the spherical
delta function can be viewed as a generalized solution of the Poisson
equation of a unit multipole charge.

The properties of generalized functions of the type of the spherical
delta function can be understood most easily with the help of Fourier
transformation. As already mentioned in Section \ref{Sec:FourierTr},
Fourier transformation is defined in the sense of classical analysis
only for functions that are absolutely integrable, i.e., which belong to
the function space $L^1 (\mathbb{R}^3)$. By means of a suitable limiting
procedure, Fourier transformation can be extended uniquely to give a
unitary map from the Hilbert space $L^2 (\mathbb{R}^3)$ of square
integrable functions onto itself (see for example \cite[Theorem IX.6 on
p.\ 10]{Reed/Simon/1975}). A further extension of Fourier transformation
to the space of tempered distributions is also possible (see for example
\cite[Theorem IX.2 on p.\ 5]{Reed/Simon/1975}).

The extensibility of Fourier transformation to nonclassical function
spaces is very important for our purposes. For example, the Coulomb
potential is neither absolutely integrable nor square integrable.
Nevertheless, it is possible to define its Fourier transform in the
sense of distributions \cite[Eq.\ (2) on p.\ 
194]{Gelfand/Vol1/Shilov/1964}:
\begin{equation}
  \label{FT_CoulombPot}
\frac{(2/\pi)^{1/2}}{p^2} \; = \; (2\pi)^{-3/2} \,
\int \, \frac{\mathrm{e}^{-\mathrm{i} \bm{p} \cdot \bm{r}}}{r} \,
\mathrm{d}^3 \bm{r} \, .
\end{equation}
This relationship makes it possible to convert multicenter integrals,
which describe the Coulomb interaction of classical or nonclassical
charge densities, into momentum space integrals according to
(\ref{Coulomb_fg}). This is a common approach in the case of
exponentially decaying basis functions (see for example
\cite{Weniger/Grotendorst/Steinborn/1986b} or
\cite{Grotendorst/Steinborn/1988} and references therein).

In the same way, we obtain the Fourier transform of the irregular solid
harmonic, which again holds in the sense of distributions \cite[Eq.\ 
(3.11)]{Weniger/Grotendorst/Steinborn/1986b}:
\begin{align}
  \label{FT_Zlm}
  \bar{\mathcal{Z}}_{\ell}^{m} (\bm{p}) & \; = \; (2\pi)^{-3/2} \, \int
  \, \mathrm{e}^{-\mathrm{i} \bm{p} \cdot \bm{r}} \,
  \mathcal{Z}_{\ell}^{m} (\bm{r}) \, \mathrm{d}^3 \bm{r}
  \notag \\
  & \; = \; (2/\pi)^{1/2} \, \frac{(-1)^{\ell}}{(2\ell-1)!!} \,
  \frac{\mathcal{Y}_{\ell}^{m} (\mathrm{i} \bm{p})}{p^2} \, .
\end{align}

It is also quite instructive to study the Fourier transforms of
distributions as for instance the three-dimensional delta function or
the spherical delta function. If we set $f (\bm{r}-\bm{r}') = \delta
(\bm{r}-\bm{r}')$ in the convolution integral (\ref{Convol_fg}), we see
that the Fourier transform of the delta function is a constant:
\begin{equation}
  \label{FT_delta}
\bar{\delta} (\bm{p}) \; = \; (2\pi)^{-3/2} \int \,
\mathrm{e}^{-\mathrm{i} \bm{p} \cdot \bm{r}} \, \delta (\bm{r})
\, \mathrm{d}^3 \bm{r} \; = \; (2\pi)^{-3/2} \, .
\end{equation}
Similarly, we find that the Fourier transform of the spherical delta
function is essentially a regular solid harmonic in momentum space:
\begin{align}
  \label{FT_delta_lm}
  \bar{\delta}_{\ell}^{m} (\bm{p}) & \; = \; (2\pi)^{-3/2} \, \int \,
  \mathrm{e}^{-\mathrm{i} \bm{p} \cdot \bm{r}} \, \delta_{\ell}^{m}
  (\bm{r}) \, \mathrm{d}^3 \bm{r}
  \notag \\
  & \; = \; (2\pi)^{-3/2} \, \frac{(-1)^{\ell}}{(2\ell-1)!!} \,
  \mathcal{Y}_{\ell}^{m} (\mathrm{i} \bm{p}) \, .
\end{align}

In physics, it is common to introduce the three-dimensional delta
function via the Poisson equation (\ref{PoissonEq_CP}). Both the Poisson
equation (\ref{PoissonEq_CP}) as well as its anisotropic generalization
(\ref{PoissonEq_Zlm}) follow from the Fourier transforms
(\ref{FT_CoulombPot}) and (\ref{FT_Zlm}), respectively. We only have to
take into account that the function $1/p^2$, which occurs in both
(\ref{FT_CoulombPot}) and (\ref{FT_Zlm}), is canceled by the Laplacian
$\nabla^2$, whose Fourier transform is $- p^2$. 

It is, however, just as well possible to introduce the there-dimensional
delta function as well as the spherical delta function via the
differential operator $1 - \alpha^{-2} \nabla^2$ of the modified
Helmholtz equation, whose Fourier transform is given by $[\alpha^2 +
p^2]/\alpha^2$. This follows at once from that fact that the Yukawa
potential \cite{Yukawa/1935}, which is an exponentially screened Coulomb
potential with screening parameter $\alpha$, is also a special $B$
function according to
\begin{equation}
  \label{YukawaPot_Bfun}
\frac{\mathrm{e}^{-\alpha r}}{r} \; = \; 
\alpha \, \hat{k}_{-1/2} (\alpha r) \; = \;
(4\pi)^{1/2} \, \alpha \,  B_{0, 0}^{0} (\alpha, \bm{r}) \, .
\end{equation}
Obviously, the Yukawa potential approaches the Coulomb potential in the
limit of vanishing screening, i.e., for $\alpha \to 0$.

If we set $n=\ell=m=0$ in (\ref{FT_B_Fun}), we essentially obtain the
Fourier transform of the Yukawa potential:
\begin{align}
  \label{FT_B000}
  \bar{B}_{0, 0}^{0} (\alpha, \bm{p}) & \; = \; (2\pi)^{-3/2} \, \int 
  \, \mathrm{e}^{- \mathrm{i} \bm{p} \cdot \bm{r}} \, B_{0,0}^{0}
  (\alpha, \bm{r}) \, \mathrm{d}^3 \bm{r}
  \notag \\
  & \; = \; (2\pi^2)^{-1/2} \, \frac{\alpha^{-1}}{[\alpha^2 + p^2]} \, .
\end{align}
Next, we use (\ref{Convol_fg}) to compute the convolution of a
relatively arbitrary function $f : \mathbb{R}^3 \to \mathbb{C}$ with the
Yukawa potential \cite[Eq.\ (6.9)]{Weniger/Steinborn/1983c}:
\begin{align}
  & \int \, B_{0, 0}^{0} (\alpha, [\bm{r}-\bm{r}']) \, f (\bm{r}') \,
  \mathrm{d} \bm{r} \; = \; \int \, \mathrm{e}^{\mathrm{i} \bm{r} \cdot
    \bm{p}} \, \bar{B}_{0, 0}^{0} (\alpha, \bm{p}) \, \bar{f} (\bm{p})
  \, \mathrm{d}^3 \bm{p}
  \notag \\
  & \qquad \quad \; = \; \frac{\alpha^{-1}}{(2\pi^2)^{1/2}} \, \int \,
  \frac{\mathrm{e}^{\mathrm{i} \bm{r} \cdot \bm{p}}}{\alpha^2 + p^2} \,
  \, \bar{f} (\bm{p}) \, \mathrm{d}^3 \bm{p} \, .
\end{align}
If we now apply the differential operator $1 - \alpha^{-2} \nabla^2$ to
the convolution integral, interchange integration and differentiation,
and use (\ref{Shift_n_Blm}), we see that $B_{-1, 0}^{0} (\alpha, \bm{r})
= [1 - \alpha^{-2} \nabla^2] B_{0, 0}^{0} (\alpha, \bm{r})$ is
proportional to the three-dimensional delta function \cite[Eq.\ 
(6.10)]{Weniger/Steinborn/1983c}:
\begin{align}
  \label{delta_B000}
  & \bigl[ 1 - \alpha^{-2} \nabla^2 \bigr] \int \, B_{0, 0}^{0} (\alpha,
  [\bm{r}-\bm{r}']) \, f (\bm{r}') \, \mathrm{d} \bm{r}
  \notag \\
  & \qquad \; = \; \int \, B_{-1, 0}^{0} (\alpha, [\bm{r}-\bm{r}']) \, f
  (\bm{r}') \, \mathrm{d} \bm{r}
  \notag \\
  & \qquad \; = \; \frac{(4\pi)^{1/2}}{\alpha^3} \, (2\pi)^{-3/2} \,
  \int \, \mathrm{e}^{\mathrm{i} \bm{r} \cdot \bm{p}} \, \, \bar{f}
  (\bm{p}) \, \mathrm{d}^3 \bm{p} \; = \;
  \frac{(4\pi)^{1/2}}{\alpha^{3}} \, f (\bm{r}) \, .
\end{align}
We thus obtain the following exponentially screened variant of the
Poisson equation (\ref{PoissonEq_CP}):
\begin{equation}
\bigl[ 1 - \alpha^{-2} \nabla^2 \bigr] \, B_{0, 0}^{0} (\alpha, \bm{r}) 
\; = \; \frac{(4\pi)^{1/2}}{\alpha^{3}} \, \delta (\bm{r}) \, .
\end{equation}
This relationship can also be expressed in terms of the Yukawa potential:
\begin{equation}
  \label{GF_ModHelmholtzEq}
\bigl[ \alpha^{2} - \nabla^2 \bigr] \, \frac{\mathrm{e}^{-\alpha r}}{r}
\; = \; 4\pi \, \delta (\bm{r}) \, .
\end{equation}
For $\alpha = 0$, we obtain the Poisson equation (\ref{PoissonEq_CP}).

The screened Poisson equation (\ref{GF_ModHelmholtzEq}) expresses the
well known fact that the Yukawa potential is the Green's function of the
modified Helmholtz equation (see for example \cite[Table 16.1 on p.\
912]{Arfken/1985}). 

An anisotropic generalization of the approach described above is also
possible. If we set $n=-\ell$ in (\ref{STGO_Bn00}), we see that the
application of the spherical tensor gradient operator to the Yukawa
potential yields the so-called modified Helmholtz harmonic \cite[Eq.\ 
(6.9)]{Weniger/Steinborn/1985}:
\begin{equation}
B_{-\ell, \ell}^{m} (\alpha, \bm{r}) \; = \; 
(4\pi)^{1/2} \, (-\alpha)^{-\ell} \, \mathcal{Y}_{\ell}^{m} (\nabla) \, 
B_{0, 0}^{0} (\alpha, \bm{r}) \, 
\end{equation}
In the limit of vanishing screening, the modified Helmholtz harmonic
approaches an irregular solid harmonic according to \cite[Eq.\ 
(3.10)]{Weniger/Grotendorst/Steinborn/1986b}
\begin{equation}
\mathcal{Z}_{\ell}^{m} (\bm{r}) \; = \; [(2\ell-1)!!]^{-1} \, 
\lim_{\alpha \to 0} \, 
\bigl[ \alpha^{\ell+1} B_{-\ell, \ell}^{m} (\alpha, \bm{}r) \bigr] \, .
\end{equation} 

If we set $n=-\ell$ in (\ref{FT_B_Fun}), we obtain the Fourier transform
of the modified Helmholtz harmonic \cite[Eq.\ 
(A.1)]{Weniger/Steinborn/1983a}:
\begin{equation}
\bar{B}_{-\ell, \ell}^{m} (\alpha, \bm{p}) \; = \; 
(2/\pi)^{1/2} \, \frac{\alpha^{-\ell-1}}{[\alpha^2 + p^2]} 
\, \mathcal{Y}_{\ell}^{m} (- i \bm{p}) \, .
\end{equation}
Both the Fourier integral producing this relationship as well as the
inverse Fourier integral representation for $B_{-\ell, \ell}^{m}
(\alpha, \bm{r})$ do not exist in the sense of classical analysis. In
\cite[Appendix]{Weniger/Steinborn/1983a} it was, however, shown that the
classically divergent Fourier integral representation for $B_{-\ell,
  \ell}^{m} (\alpha, \bm{r})$ can be regularized by employing a suitable
rational cutoff function.

If we now proceed as in the case of (\ref{delta_B000}) and also use
(\ref{YlmNabla_PlaneWave}), we see that $B_{-\ell-1, \ell}^{m} (\alpha,
\bm{r}) = [1 - \alpha^{-2} \nabla^2] B_{-\ell, \ell}^{m} (\alpha,
\bm{r})$ is proportional to the spherical delta function \cite[Eq.\ 
(6.17) - (6.19)]{Weniger/Steinborn/1983c}:
\begin{align}
  \label{delta_lm_Bnlm}
  & \bigl[ 1 - \alpha^{-2} \nabla^2 \bigr] \, \int \, B_{-\ell,
    \ell}^{\ell} (\alpha, [\bm{r}-\bm{r}']) \, f (\bm{r}') \, \mathrm{d}
  \bm{r}
  \notag \\
  & \qquad \; = \; \int \, B_{-\ell-1, \ell}^{\ell} (\alpha,
  [\bm{r}-\bm{r}']) \, f (\bm{r}') \, \mathrm{d} \bm{r}
  \notag \\
  & \qquad \; = \; \frac{(2/\pi)^{1/2}}{\alpha^{\ell+3}} \, \int \,
  \mathrm{e}^{\mathrm{i} \bm{r} \cdot \bm{p}} \, \mathcal{Y}_{\ell}^{m}
  (- i \bm{p}) \, \, \bar{f} (\bm{p}) \, \mathrm{d}^3 \bm{p}
\notag \\
& \qquad 
  \; = \; (-1)^{\ell} \, \frac{4\pi}{\alpha^{\ell+3}} \, 
  \mathcal{Y}_{\ell}^{m} (\nabla) \, f (\bm{r}) \, .
\end{align}
We thus obtain the following exponentially screened variant of the
anisotropic Poisson equation (\ref{PoissonEq_Zlm}):
\begin{align}
  & \bigl[ 1 - \alpha^{-2} \nabla^2 \bigr] \, B_{-\ell, \ell}^{m}
  (\alpha, \bm{r})
  \notag \\
  & \qquad \; = \; (-1)^{\ell} \, \frac{4\pi}{\alpha^{\ell+3}} \,
  \mathcal{Y}_{\ell}^{m} (\nabla) \, \delta (\bm{r}) \; = \;
  \frac{4\pi}{\alpha^{\ell+3}} \, (2\ell-1)!! \, 
  \delta_{\ell}^{m} (\bm{r}) \, .
\end{align}
In view of this relationship and also because of (\ref{Shift_n_Blm}) it
makes sense to define a distributional $B$ function as the following
derivative of the three-dimensional Dirac delta function \cite[Eq.\ 
(6.20)]{Weniger/Steinborn/1983c}:
\begin{equation}
  \label{Def_Distrib_B_Fun}
B_{-k-\ell,\ell}^{m} (\alpha, \bm{r}) \; = \; 
\frac{(2\ell-1)!! \, 4\pi}{\alpha^{\ell+3}} \, 
\left[1 - \alpha^{-2} \nabla^2 \right]^{k-1} \,
\delta_{\ell}^{m} (\bm{r}) \, , \quad k \in \mathbb{N} \, .
\end{equation}
The fact that the distributional $B$ function $B_{-\ell-1, \ell}^{B}$ is
proportional to the spherical delta function can also be seen by setting
$n_1=-\ell_1-1$ in the convolution integral (\ref{Conv_Bnlm_alpha}).
Then, we obtain apart from a different prefactor the expression
(\ref{STGO_Bnlm}) for the product $\mathcal{Y}_{\ell_1}^{m_1} (\nabla)
\, B_{n_2,\ell_2}^{m_2} (\alpha, \bm{r})$.

The distributional nature of some $B$ functions follows also from the
fact that the Fourier transforms (\ref{FT_B_Fun}) of $B$ functions
satisfy for all $n \in \mathbb{Z}$ and for all $n \in \mathbb{N}_0$ the
following functional equations \cite[Eqs.\ (6.21) -
(6.23)]{Weniger/Steinborn/1983c}:
\begin{subequations}
  \label{FuncEqs_FT_Bnlm}
\begin{align}
  \bar{B}_{n,\ell}^{m} (\alpha, \bm{p}) & \; = \;
  [\alpha^2/(\alpha^2+p^2)] \, \bar{B}_{n-1,\ell}^{m} (\alpha, \bm{p})
  \, ,
  \\
  \bar{B}_{n,\ell}^{m} (\alpha, \bm{p}) & \; = \;
  [(4\pi)^{1/2}/\alpha^{\ell}] \, \mathcal{Y}_{\ell}^{m} (-\mathrm{i}
  \bm{p}) \, \bar{B}_{n+\ell, 0}^{0} (\alpha, \bm{p}) \, ,
  \\
  \bar{B}_{-1,0}^{0} (\alpha, \bm{p}) & \; = \; \alpha^{-3} \,
  (2\pi^2)^{-1/2} \, .
\end{align}
\end{subequations}
These functional equations in momentum space show that there is an
intimate relationship between $B$ functions, the differential operator
of the modified Helmholtz equation, the spherical tensor gradient
operator, and the three-dimensional delta function.

There is a simple and intuitive interpretation of distributional $B$
functions. Because of the factorial $(n+\ell)!$ in the denominator on
the right-hand side of (\ref{Def:B_Fun}), $B$ functions are defined in
the sense of classical analysis only if $n+\ell \ge 0$ holds. However,
the definition of a $B$ function remains meaningful even for $n+\ell <
0$. If $\bm{r} \neq \bm{0}$, the value of $B_{-k-\ell, \ell}^{m}
(\bm{r})$ with $k \in \mathbb{N}$ is because of the singular factorial
$(-k)! = \Gamma (-k+1)$ zero, but for $\bm{r} = \bm{0}$, its value is
$\infty/\infty$ and therefore undefined. Thus, the mathematical nature
of a $B$ function with $n+\ell < 0$ as well as its value for $\bm{r} =
\bm{0}$ has to be analyzed with the help of suitable limiting
procedures.

\typeout{==> Section: Addition Theorems}
\section{Addition Theorems}
\label{Sec:AdditionThm}

In many subfields of physics and physical chemistry -- for example in
electrodynamics \cite{Jackson/1975}, in classical field theory
\cite{Jones/1985}, or in the theory of intermolecular forces
\cite{Stone/1996} -- an essential step towards a solution of the problem
under consideration consists in a separation of variables.

Principal tools, which can accomplish such a separation of variables,
are so-called addition theorems. These are expansions of a given
function $f (\bm{r} \pm \bm{r}')$ with $\bm{r}, \bm{r}' \in
\mathbb{R}^3$ in products of other functions that only depend on either
$\bm{r}$ or $\bm{r}'$.

In view of the importance of addition theorems, it is not surprising
that there is a very extensive literature. Consequently, any attempt to
provide a reasonably complete bibliography would clearly be beyond the
scope of this article. The interested reader is referred to the long,
but nevertheless incomplete lists of references in
\cite{Weniger/2000a,Weniger/2002}.

Addition theorems have played a major role in my own research. I first
applied addition theorems for the evaluation of some multicenter
integrals of $B$ functions in my Master's thesis \cite{Weniger/1977},
which was published in condensed form in \cite{Steinborn/Weniger/1977}.
In my PhD thesis \cite{Weniger/1982} and also afterwards, I preferred
Fourier transformation for the evaluation of multicenter integrals of
$B$ functions, but I later worked on the derivation of addition theorems
\cite{Homeier/Weniger/Steinborn/1992a,Weniger/1985,Weniger/2000a,%
Weniger/2002,Weniger/Steinborn/1985,Weniger/Steinborn/1989b}.

In atomic or molecular calculations, we are predominantly interested in
irreducible spherical tensors of the type of
(\ref{Def_IrrSphericalTensor}). Moreover, the convenient orthonormality
of the spherical harmonics makes it highly desirable that the functions
of either $\bm{r}$ or $\bm{r}'$, which occur in the expansion of $f
(\bm{r} \pm \bm{r}')$, are also irreducible spherical tensors. Thus, the
addition theorems, we are interested in, are expansions in terms of
spherical harmonics with arguments $\theta, \phi = \bm{r}/r$ and
$\theta', \phi' = \bm{r}'/r'$, respectively.

The best known example of such an addition theorem is the Laplace
expansion of the Coulomb potential in terms of spherical harmonics:
\begin{align}
  \label{LapExp}
  \frac{1}{\vert {\bm{r}} \pm {\bm{r}}' \vert} & \; = \; 4\pi \,
  \sum_{\lambda=0}^{\infty} \, \frac{(\mp 1)^{\lambda}}{2\lambda+1} \,
  \sum_{\mu=-\lambda}^{\lambda} \, \bigl[ \mathcal{Y}_{\lambda}^{\mu}
  ({\bm{r}_{<}}) \bigr]^{*} \, \mathcal{Z}_{\lambda}^{\mu}
  ({\bm{r}}_{>}) \, ,
  \notag \\
  & \qquad r_{<} = \min (r,r')\, , \quad r_{>} = \max (r,r') \, .
\end{align}
The Laplace expansion leads to a separation of the variables ${\bm{r}}$
and ${\bm{r}}'$. However, its right-hand side depends on ${\bm{r}}$ and
${\bm{r}}'$ only indirectly via the vectors $\bm{r}_{<}$ and
$\bm{r}_{>}$ which satisfy $\vert \bm{r}_{<} \vert < \vert \bm{r}_{>}
\vert$. Hence, the Laplace expansion has a two-range form, depending on
the relative size of $\bm{r}$ and $\bm{r}'$. This is a complication
which occurs frequently among addition theorems. As discussed in more
details in \cite{Weniger/2000a,Weniger/2002}, addition theorems have a
two-range form if they are pointwise convergent three-dimensional Taylor
expansions and if the function $f (\bm{r} \pm \bm{r}')$, which is to be
expanded, is not analytic at the origin.

The undeniably troublesome two-range form of an addition theorem can be
avoided if $f : \mathbb{R}^3 \to \mathbb{C}$ belongs to the Hilbert
space $L^2 (\mathbb{R}^3)$ of square integrable functions or to other,
closely related function spaces as for example Sobolev spaces that are
proper subspaces of $L^2 (\mathbb{R}^3)$ (compare for instance
\cite{Filter/Steinborn/1980,Homeier/Weniger/Steinborn/1992a,%
  Weniger/1985,Weniger/Steinborn/1984} and references therein). For the
sake of simplicity, let us assume that a discrete function set $\{
\Psi_{n, \ell}^{m} (\bm{r}) \}_{n, \ell, m}$ is complete and orthonormal
in the Hilbert space $L^2 (\mathbb{R}^3)$. Then, an addition theorem for
$f (\bm{r} \pm \bm{r}')$, which converges in the mean with respect to
the norm of $L^2 (\mathbb{R}^3)$, can be derived by expanding $f$ in
terms of the functions $\{ \Psi_{n, \ell}^{m} (\bm{r}) \}_{n, \ell, m}$:
\begin{subequations}
   \label{OneRangeAddTheor}
\begin{align}
  f (\bm{r} \pm \bm{r}') & \; = \; \sum_{n \ell m} \, C_{n, \ell}^{m}
  (f; \bm{r}') \, \Psi_{n, \ell}^{m} (\bm{r}) \, ,
  \\
  C_{n, \ell}^{m} (f; \bm{r}') & \; = \; \int \, \bigl[ \Psi_{n,
    \ell}^{m} (\bm{r}) \bigr]^{*} \, f (\bm{r} \pm \bm{r}') \,
  \mathrm{d}^3 \bm{r} \, .
\end{align}
\end{subequations}
Such an expansion is a one-range addition theorem since the dependence
on $\bm{r}$ is entirely contained in the functions $\Psi_{n, \ell}^{m}
(\bm{r})$, whereas $\bm{r}'$ occurs only in the expansion coefficients
$C_{n, \ell}^{m} (f; \bm{r}')$ which are overlap integrals. With minimal
modifications, this approach works also if the functions $\{ \Psi_{n,
  \ell}^{m} (\bm{r}) \}_{n, \ell, m}$ are complete and orthonormal in a
suitable Sobolev space.

At first sight, it looks as if one-range addition theorems of the type
of (\ref{OneRangeAddTheor}) are clearly superior to two-range addition
theorems, but a balanced assessment of their relative merits is not so
easy. Firstly, one-range addition theorems usually have a more
complicated structure than two-range addition theorems (typically,
one-range addition theorems contain one additional infinite summation).
Secondly, one-range addition theorems normally converge only in the mean
with respect to the norm of the Hilbert space $L^2 (\mathbb{R}^3)$, but
not necessarily pointwise. In some cases, this can lead to convergence
problems. The probably the most severe disadvantage is that the approach
sketched above works only if $f$ is an element of a suitable Hilbert or
Sobolev space. Many functions of considerable practical importance such
as the Coulomb potential or the irregular solid harmonic do not belong
to $L^2 (\mathbb{R}^3)$, let alone to a suitable Sobolev space.
Therefore, it is not possible to derive one-range addition theorems by
expanding them in terms of function sets that are complete and
orthonormal with respect to a scalar product that involves an
integration over the whole $\mathbb{R}^3$.

In this article, I will only discuss addition theorems that converge
pointwise, i.e., which can be viewed to be three-dimensional Taylor
expansions. Moreover, I will concentrate on addition theorems that can
be derived with the help of the spherical tensor gradient operator
${\mathcal{Y}}_{\ell}^{m} (\nabla)$.

As discussed in Section \ref{Sec:HobsonDiffTheor}, the irregular solid
harmonic $\mathcal{Z}_{\ell}^{m} (\bm{r})$ can according to
(\ref{YllNabla_Coul}) be obtained by applying ${\mathcal{Y}}_{\ell}^{m}
(\nabla)$ to the Coulomb potential. Thus, it should be possible to
derive the addition theorem of the irregular solid harmonic by applying
either ${\mathcal{Y}}_{\ell}^{m} (\nabla_{<})$ or
${\mathcal{Y}}_{\ell}^{m} (\nabla_{>})$ to the Laplace expansion
(\ref{LapExp}). As shown in \cite[Section IV]{Weniger/Steinborn/1985},
this is indeed possible. Moreover, it was shown in \cite[Sections V and
VI]{Weniger/Steinborn/1985} that the addition theorems of the Helmholtz
and the modified Helmholtz harmonics can also be derived by applying
either ${\mathcal{Y}}_{\ell}^{m} (\nabla_{<})$ or
${\mathcal{Y}}_{\ell}^{m} (\nabla_{>})$ to the simpler addition theorems
of the corresponding scalar functions. It is also possible to derive the
addition theorem of $B$ functions by applying the spherical tensor
gradient operator to the addition theorem of the reduced Bessel
functions \cite[Section V]{Weniger/Steinborn/1989b}.

The approach described above requires that for a given irreducible
tensor a scalar function can be found which satisfies
(\ref{Def_Phi_l2m2}) and which possesses a suitable addition theorem.
This obviously limits the practical usefulness of this approach. It is,
however, possible to derive addition theorems of essentially arbitrary
irreducible spherical tensors from the scratch with the help of the
spherical tensor gradient operator.

As is well known, addition theorems can formally be obtained by doing
three-dimensional Taylor expansions (see for example \cite[p.\ 
181]{Biedenharn/Louck/1981a}):
\begin{equation}
  \label{ExpDifOp}
f (\bm{r} + \bm{r}') \; = \; \sum_{n=0}^{\infty} \,
\frac{(\bm{r}' \cdot \nabla)^n}{n!} \, f (\bm{r})
\; = \;
\mathrm{e}^{\bm{r}' \cdot \nabla} \, f (\bm{r}) \, .
\end{equation}
Thus, the translation operator 
\begin{equation}
  \label{CartTransOp}
\mathrm{e}^{\bm{r}' \cdot \nabla} \; = \; 
\mathrm{e}^{x' \partial /\partial x} \,
\mathrm{e}^{y' \partial /\partial y} \,
\mathrm{e}^{z' \partial /\partial z}
\end{equation}
generates $f (\bm{r} + \bm{r}')$ by doing a three-dimensional Taylor
expansion of $f$ around $\bm{r}$ with shift vector $\bm{r}'$. Moreover,
the variables $\bm{r}$ and $\bm{r}'$ are separated. Thus, the series
expansion (\ref{ExpDifOp}) is indeed an addition theorem.

We could also expand $f$ around $\bm{r}'$ and use $\bm{r}$ as the shift
vector. This would produce an addition theorem for $f (\bm{r} +
\bm{r}')$ in which the roles of $\bm{r}$ and $\bm{r}'$ are interchanged.
Both approaches are mathematically legitimate and equivalent if $f$ is
analytic at $\bm{r}$, $\bm{r}'$, and $\bm{r} + \bm{r}'$ for essentially
arbitrary vectors $\bm{r}, \bm{r}' \in \mathbb{R}^3$. Unfortunately,
this is normally not true. Most functions, that are of interest in the
context of atomic and molecular quantum mechanics, are either singular
at the origin or are not analytic at the origin. Obvious examples are
the Coulomb potential, which is singular at the origin, or the $1 s$
hydrogen eigenfunction, which possesses a cusp at the origin. In fact,
all the commonly used exponentially decaying functions as for example
Slater-type functions or also $B$ functions are not analytic at the
origin. 

The reason for the non-analyticity is that the three-dimensional
distance $r = \bigl[ x^2 + y^2 + z^2 \bigr]^{1/2}$ is not analytic with
respect to $x$, $y$, and $z$ at the origin $\hm{r} = \hm{0}$. This
implies that all odd powers $r^{2n+1}$ with $n \in \mathbb{N}_0$ are
also not analytic at the origin (compare also the discussion related to
\cite[Eq.\ (5.9)]{Weniger/1985}). In contrast, $r^2 = x^2 + y^2 + z^2$
and the regular solid harmonic $\mathcal{Y}_{\ell}^{m} (\bm{r})$ are
analytic since they are polynomials in $x$, $y$, and $z$. Consequently,
a $1 s$ Gaussian function $\exp (- \alpha r^2)$, which possesses an
expansion in even powers $r^{2n}$, is analytic at $\bm{r} = \bm{0}$,
but a $1 s$ Slater-type function $\exp (- \alpha r)$ is not.

Thus, for the derivation of addition theorems for functions, that are
not analytic at the origin, we have to use the translation operator in
the following form,
\begin{equation}
  \label{TransOp<>}
f (\bm{r}_{<} + \bm{r}_{>}) \; = \;
\sum_{n=0}^{\infty} \,
\frac{(\bm{r}_{<} \cdot \nabla_{>})^n}{n!} \,
f (\bm{r}_{>}) \; = \;
\mathrm{e}^{\bm{r}_{<} \cdot \nabla_{>}} \,
f (\bm{r}_{>}) \, ,
\end{equation}
where $\vert \bm{r}_{<} \vert < \vert \bm{r}_{>} \vert$. In this way,
the convergence of the three-dimensional Taylor expansion is guaranteed
provided that $f$ is analytic everywhere except possibly at the origin.
Thus, the non-analyticity of $B$ functions and of all the other commonly
occurring exponentially decaying functions at the origin has a
far-reaching consequence: Their pointwise convergent addition theorems
must have a two-range form in order to guarantee convergence (see also
\cite{Weniger/2002} and references therein).

From a practical point of view, the translation operator
$\mathrm{e}^{\bm{r}_{<} \cdot \nabla_{>}}$ in its Cartesian form does
not seem to be a particularly useful analytical tool. We are usually
interested in addition theorems of irreducible spherical tensors, which
are defined in terms of the spherical polar coordinates $r$, $\theta$,
and $\phi$. Differentiating them with respect to the Cartesian
components $x$, $y$, and $z$ would lead to extremely messy expressions
and to difficult technical problems. Thus, it is tempting, but
nevertheless superficial to conclude that the translation operator
$\mathrm{e}^{\bm{r}_{<} \cdot \nabla_{>}}$ provides only a formal
solution to the problem of separating the variables $\bm{r}$ and
$\bm{r}'$ of a function $f (\bm{r} + \bm{r}')$.

The crucial step, which ultimately makes the Taylor expansion method
practically useful, is the expansion of the translation operator
$\mathrm{e}^{\bm{r}_{<} \cdot \nabla_{>}}$ in terms of differential
operators that are irreducible spherical tensors. The starting point is
an expansion of $\exp (\bm{a} \cdot \bm{b})$ with $\bm{a}, \bm{b} \in
\mathbb{R}^3$ in terms of modified Bessel functions and Legendre
polynomials \cite[p.\ 108]{Magnus/Oberhettinger/Soni/1966}:
\begin{equation}
  \label{I_Pl_exp}
\mathrm{e}^{\bm{a} \cdot \bm{b}} \; = \;
\mathrm{e}^{a b \cos \theta} \; = \;
\left( \frac{\pi}{2 a b} \right)^{1/2} \, \sum_{\ell=0}^{\infty} \,
(2\ell+1) \, I_{\ell+1/2} (ab) \, P_{\ell} (\cos \theta) \, .
\end{equation}
Next, the series expansion for the modified Bessel function
$I_{\ell+1/2}$ \cite[p.\ 66]{Magnus/Oberhettinger/Soni/1966} is inserted
into (\ref{I_Pl_exp}), and spherical harmonics are introduced with the
help of the so-called addition theorem of the Legendre polynomials (see
for example \cite[p.\ 303]{Biedenharn/Louck/1981a}). This yields the
following expansion of $\mathrm{e}^{\bm{a} \cdot \bm{b}}$ in terms of
regular solid harmonics and even powers of the vectors $\bm{a}$ and
$\bm{b}$:
\begin{equation}
  \label{ST_Exp_e_ab}
\mathrm{e}^{\bm{a} \cdot \bm{b}} \; = \;
2 \pi \, \sum_{\ell=0}^{\infty} \, \sum_{m=-\ell}^{\ell} \,
\left[ \mathcal{Y}_{\ell}^{m} (\bm{a}) \right]^{*} \,
\mathcal{Y}_{\ell}^{m} (\bm{b}) \, \sum_{k=0}^{\infty} \,
\frac{\bm{a}^{2k} \, \bm{b}^{2k}}
{2^{\ell+2k} k! (1/2)_{\ell+k+1}} \, .
\end{equation}
The powers $\bm{a}^{2k}$ and $\bm{b}^{2k}$ are irreducible spherical
tensors of rank zero, and the solid harmonics are tensors of rank
$\ell$.

The expansion (\ref{ST_Exp_e_ab}) is obtained from (\ref{I_Pl_exp}) by
rearranging the Cartesian components of the vectors $\bm{a}$ and
$\bm{b}$. Accordingly, it holds for essentially arbitrary vectors
$\bm{a}, \bm{b} \in \mathbb{R}^3$, and we can choose $\bm{a} =
\bm{r}_{<}$ and $\bm{b} = \nabla_{>}$:
\begin{equation}
  \label{ST_TransOp}
\mathrm{e}^{\bm{r}_{<} \cdot \nabla_{>}} \; = \; 
2 \pi \, \sum_{\ell=0}^{\infty} \, \sum_{m=-\ell}^{\ell} \,
\left[ \mathcal{Y}_{\ell}^{m} (\bm{r}_{<}) \right]^{*} \,
\mathcal{Y}_{\ell}^{m} (\nabla_{>}) \, \sum_{k=0}^{\infty} \,
\frac{\bm{r}_{<}^{2k} \, \nabla_{>}^{2k}}
{2^{\ell+2k} k! (1/2)_{\ell+k+1}} \, .
\end{equation}
It seems that this expansion was first published by Santos \cite[Eq.\ 
(A.6)]{Santos/1973}, who emphasized that this expansion should be useful
for the derivation of addition theorems, but apparently never used it
for that purpose.

It follows from the expression (\ref{Def_Lapl_r_theta_phi}) of the
Laplacian $\nabla^2$ in spherical polar coordinates and from the
tensorial nature of the spherical tensor gradient operator (compare
(\ref{YlmNab2Flm_GenStruc}) and the numerous expressions for $\gamma
_{\ell_1 \ell_2}^{\ell} (r)$ given in Section \ref{Sec:FourierTr}) that
in (\ref{ST_TransOp}) we only have to do differentiations with respect
to the radial variable $r_{>}$. This greatly simplifies practical
applications. In \cite{Weniger/2000a}, it was shown that starting from
the expansion (\ref{ST_TransOp}) the addition theorems of the Coulomb
potential, the regular and irregular solid harmonics, and the Yukawa
potential can be derived quite easily, and in \cite{Weniger/2002},
several different forms of the addition theorem of the $B$ functions was
derived in this way.

As a further demonstration of the usefulness of the expansion
(\ref{ST_TransOp}) of the translation operator in terms of irreducible
spherical tensors I will now derive an addition theorem of the function
$r^{\nu} {\mathcal{Y}_{\ell}^{m}} (\bm{r})$ with $\nu \in \mathbb{R}$.

Our starting point the the following relationship which follows at once
from (\ref{YlmNabla_phi}):
\begin{equation}
r^{\nu} \, {\mathcal{Y}_{\ell}^{m}} (\bm{r}) \; = \; 
\frac{1}{2^{\ell} (1+\nu/2)_{\ell}} \, \mathcal{Y}_{\ell}^{m} (\nabla) 
\, r^{\nu+2\ell} \, .
\end{equation}
If we combine this with (\ref{ST_TransOp}) and linearize the spherical
tensor gradient operators according to (\ref{YlmNabla_lin}), we obtain:
\begin{align}
  \label{AT_r^ny_Ylm_1}
  & \vert \bm{r}_{<} + \bm{r}_{>} \vert^{\nu} \,
  {\mathcal{Y}_{\ell}^{m}} ( \bm{r}_{<} + \bm{r}_{>} ) \; = \;
  \frac{2\pi}{2^{\ell} (1+\nu/2)_{\ell}} \, \sum_{\ell_1=0}^{\infty} \,
  \sum_{m_1=-\ell_1}^{\ell_1} \bigl[ {\mathcal{Y}_{\ell_1}^{m_1}}
  (\bm{r}_{<}) \bigr]^{*}
  \notag \\
  & \qquad \times
  \sum_{\ell_2=\ell_2^{\mathrm{min}}}^{\ell_2^{\mathrm{max}}} \!
  {}^{(2)} \, \langle \ell_2 m+m_1 \vert \ell_1 m_1 \vert \ell m \rangle
  \, \sum_{k=0}^{\infty} \, \frac{r_{<}^{2k}}{2^{\ell_1+2k} \, k! \,
    (1/2)_{\ell_1+k+1}}
  \notag \\
  & \qquad \qquad \times \nabla_{>}^{2k+2\Delta\ell_2} \,
  \mathcal{Y}_{\ell_2}^{m+m_1} (\nabla_{>}) \, r_{>}^{\nu+2\ell} \, .
\end{align}
The abbreviations $\Delta \ell$, $\Delta \ell_1$, and $\Delta \ell_2$,
which will be used in the following formulas and which are in all cases
either zero or a positive integer, are defined by (\ref{Def_Del_l}) --
(\ref{Def_Del_l_2}),

In the next step, we use (\ref{ST_TransOp}) once more to obtain
\begin{equation}
\mathcal{Y}_{\ell_2}^{m+m_1} (\nabla) \, r^{\nu+2\ell} \; = \; 
(-2)^{\ell_2} \, (-\ell-\nu/2)_{\ell_2} \,
r^{\nu+2\ell-2\ell_2} \, {\mathcal{Y}_{\ell_2}^{m+m_1}} (\bm{r}) \, .
\end{equation}
Inserting this into (\ref{AT_r^ny_Ylm_1}) yields:
{\allowdisplaybreaks
\begin{align}
  \label{AT_r^ny_Ylm_2}
  & \vert \bm{r}_{<} + \bm{r}_{>} \vert^{\nu} \,
  {\mathcal{Y}_{\ell}^{m}} ( \bm{r}_{<} + \bm{r}_{>} ) \; = \;
  \frac{2\pi}{2^{\ell} (1+\nu/2)_{\ell}} \, \sum_{\ell_1=0}^{\infty} \,
  \sum_{m_1=-\ell_1}^{\ell_1} \bigl[ {\mathcal{Y}_{\ell_1}^{m_1}}
  (\bm{r}_{<}) \bigr]^{*}
  \notag \\
  & \qquad \times
  \sum_{\ell_2=\ell_2^{\mathrm{min}}}^{\ell_2^{\mathrm{max}}} \!
  {}^{(2)} \, (-1)^{\ell_2} \, \, \langle \ell_2 m+m_1 \vert \ell_1 m_1
  \vert \ell m \rangle
  \notag \\
  & \qquad \qquad \times \sum_{k=0}^{\infty} \,
  \frac{2^{\ell_2-\ell_1-2k} \, (-\ell-\nu/2)_{\ell_2} \, r_{<}^{2k}}
  {k! \, (1/2)_{\ell_1+k+1}}
  \notag \\
  & \qquad \qquad \qquad \times \nabla_{>}^{2k+2\Delta\ell_2} \,
  r_{>}^{\nu+2\ell-2\ell_2} \, \mathcal{Y}_{\ell_2}^{m+m_1} (\bm{r}_{>})
  \, .
\end{align}
}

Now, the only thing, that remains to be done, is the application of the
powers $\nabla_{>}^{2k+2\Delta\ell_2}$ of the Laplacian to
$r_{>}^{\nu+2\ell-2\ell_2} \, \mathcal{Y}_{\ell_2}^{m+m_1}
(\bm{r}_{>})$. For that purpose, we use the following expression which
can be deduced easily from (\ref{Def_Lapl_r_theta_phi}):
\begin{equation}
\nabla^2 \psi (r) \, \mathcal{Y}_{\ell}^{m} (\bm{r}) \; = \; 
\mathcal{Y}_{\ell}^{m} (\bm{r}) \, 
\left[ \frac{\partial^2}{\partial r^2} \, + \, \frac{2\ell+2}{r} \, 
\frac{\partial}{\partial r} \right] \, \psi (r) \, .
\end{equation}
Here, $\psi (r)$ is a scalar function. If we set $\psi (r) = r^{\sigma}$
with $\sigma \in \mathbb{R}$, we find:
\begin{equation}
\nabla^2 r^{\sigma} \, \mathcal{Y}_{\ell}^{m} (\bm{r}) \; = \;
4 \, (-\sigma/2) \, (-[\sigma+2\ell+1]/2) \, r^{\sigma-2} \, 
\mathcal{Y}_{\ell}^{m} (\bm{r}) \, .
\end{equation}
Iterating this relationship $n$ times yields:
\begin{equation}
\nabla^{2n} r^{\sigma} \, \mathcal{Y}_{\ell}^{m} (\bm{r}) \; = \;
4^n \, (-\sigma/2)_n \, (-[\sigma+2\ell+1]/2)_n \, r^{\sigma-2n} \, 
\mathcal{Y}_{\ell}^{m} (\bm{r}) \, .
\end{equation}
Thus, we obtain for the remaining differentiations:
\begin{align}
  & \nabla_{>}^{2k+2\Delta\ell_2} \, r_{>}^{\nu+2\ell-2\ell_2} \,
  \mathcal{Y}_{\ell_2}^{m+m_1} (\bm{r}_{>}) \; = \; 4^{k+\Delta \ell_2}
  \notag \\[1.5\jot]
  & \quad \times \left( - \frac{\nu+2\ell-2\ell_2}{2} \right)_{\Delta
    \ell_2} \, \left( - \frac{\nu+2\ell+1}{2} \right)_{\Delta \ell_2} \,
  \left( \frac{2\Delta\ell-\nu}{2} \right)_{k}
  \notag \\[1.5\jot]
  & \quad \quad \times \left( - \frac{2\Delta\ell_1+\nu+1}{2}
  \right)_{k} \, r_{>}^{\nu+2\Delta\ell_1-2k+1} \,
  \mathcal{Z}_{\ell_2}^{m+m_1} (\bm{r}_{>}) \, .
\end{align}
With the help of some essentially straightforward algebra, it can be
shown that the $k$ summation in (\ref{AT_r^ny_Ylm_2}) can be expression
by a Gaussian hypergeometric series ${}_2 F_1$ (for its definition, see
for instance \cite[p.\ 37]{Magnus/Oberhettinger/Soni/1966}), and we
finally obtain the following addition theorem: {\allowdisplaybreaks
\begin{align}
  \label{AddThm_r^ny_Ylm}
  & \vert \bm{r}_{<} + \bm{r}_{>} \vert^{\nu} \,
  {\mathcal{Y}_{\ell}^{m}} ( \bm{r}_{<} + \bm{r}_{>} ) \; = \;
  \frac{4\pi}{(1+\nu/2)_{\ell}} \, \sum_{\ell_1=0}^{\infty} \,
  \sum_{m_1=-\ell_1}^{\ell_1} \bigl[ {\mathcal{Y}_{\ell_1}^{m_1}}
  (\bm{r}_{<}) \bigr]^{*}
  \notag \\
  & \qquad \times
  \sum_{\ell_2=\ell_2^{\mathrm{min}}}^{\ell_2^{\mathrm{max}}} \!
  {}^{(2)} \, (-1)^{\ell_2} \, \, \langle \ell_2 m+m_1 \vert \ell_1 m_1
  \vert \ell m \rangle
  \notag \\
  & \qquad \qquad \times \frac {(-\ell-\nu/2)_{\ell_2}} {(3/2)_{\ell_1}}
  \, \left( \frac{\nu-2\Delta \ell+2}{2} \right)_{\Delta \ell_2} \,
  \left( \frac{\nu-2\Delta \ell+3}{2} \right)_{\Delta \ell_2}
  \notag \\[1.8\jot]
  & \qquad \qquad \qquad \times {}_2 F_1 \left( \frac{2\Delta
      \ell-\nu}{2}, \frac{-2\Delta\ell_1-\nu-1}{2}; \frac{2\ell_1+3}{2};
    \frac{r_{<}^{2}}{r_{>}^{2}} \right)
  \notag \\[1.8\jot]
  & \qquad \qquad \qquad \qquad \times r_{>}^{\nu+2\Delta \ell_1+1} \,
  \mathcal{Z}_{\ell_2}^{m+m_1} (\bm{r}_{>}) \, .
\end{align}
}

The addition theorem (\ref{AddThm_r^ny_Ylm}) contains many simpler
addition theorems as special cases. For example, if we set in
(\ref{AddThm_r^ny_Ylm}) $\ell=m=0$, we obtain the addition theorem of
the corresponding scalar function: 
\begin{align}
  \label{AddThm_r^ny}
  \vert \bm{r}_{<} + \bm{r}_{>} \vert^{\nu} & \; = \; 4\pi \,
  r_{>}^{\nu+1} \sum_{\lambda=0}^{\infty} \, (-1)^{\lambda} \, \frac
  {(-\nu/2)_{\lambda}}{(3/2)_{\lambda}}
  \notag \\
  & \qquad \times {}_2 F_1 \left( \frac{2\lambda-\nu}{2}, \frac{-\nu-1}{2};
    \frac{2\lambda+3}{2}; \frac{r_{<}^{2}}{r_{>}^{2}} \right)
  \notag \\
  & \qquad \qquad \times \sum_{\mu=-\lambda}^{\lambda} \bigl[
  {\mathcal{Y}_{\lambda}^{\mu}} (\bm{r}_{<}) \bigr]^{*} \,
  \mathcal{Z}_{\lambda}^{\mu} (\bm{r}_{>}) \, .
\end{align}
Of course, we could also go the other way round: We could proceed as in
\cite{Weniger/Steinborn/1985} and construct the addition theorem
(\ref{AddThm_r^ny_Ylm}) by applying either ${\mathcal{Y}}_{\ell}^{m}
(\nabla_{<})$ or ${\mathcal{Y}}_{\ell}^{m} (\nabla_{>})$ to
(\ref{AddThm_r^ny}).

If we set in (\ref{AddThm_r^ny}) $\nu=-1$, we only need
$(1/2)_{\lambda}/(3/2)_{\lambda} = 1/(2\lambda+1)$ as well as ${}_2 F_1
(\lambda+1/2, 0; \lambda+3/2; r_{<}^{2}/r_{>}^{2}) = 1$ to obtain the
Laplace expansion (\ref{LapExp}) of the Coulomb potential.

If we set in (\ref{AddThm_r^ny}) $\nu = 2n$ with $n \in \mathbb{N}_0$,
the $\lambda$ summation terminates after $\lambda = n$ because of the
Pochhammer symbol $(-n)_{\lambda}$. In addition, the hypergeometric
series ${}_2 F_1$ in (\ref{AddThm_r^ny}) terminates to become a
polynomial of degree $n-\lambda$ in $r_{<}^{2}/r_{>}^{2}$. Moreover,
$r^{2n}$ is a polynomial in $x$, $y$, $z$ and thus analytic.
Accordingly, a distinction between $\bm{r}_{<}$ and $\bm{r}_{>}$ is not
necessary and the addition theorem has a one-range form.

A detailed analysis of all special cases of the addition theorem
(\ref{AddThm_r^ny_Ylm}) would clearly be beyond the scope of this
article (see also \cite[pp.\ 
168-169]{Varshalovich/Moskalev/Khersonskii/1988}). Here, one must not
forget that the emphasis of this Section is not on the derivation of the
addition theorem (\ref{AddThm_r^ny_Ylm}). Rather, it is my hope that the
fairly effortless derivation of this addition theorem convinces the
reader that the expansion (\ref{ST_TransOp}) of the translation operator
in terms of irreducible spherical tensors is indeed a highly useful
analytical tool. Of course, a sceptical reader may well argue that it is
unlikely that convenient explicit expression for the necessary radial
differentiations of arbitrary order can always be found, as it was the
case with (\ref{AddThm_r^ny_Ylm}). This certainly true. But even if we
can only do the differentiations explicitly up to a finite maximum
order, the tensorial expansion (\ref{ST_TransOp}) permits at least the
construction of approximations to addition theorems. Computer algebra
systems like Maple or Mathematica should be helpful in this respect.

\typeout{==> Section: Summary and Conclusions}
\section{Summary and Conclusions}
\label{Sec:SummConcl}

The regular solid harmonic ${\mathcal{Y}}_{\ell}^{m} (\bm{r}) = r^{\ell}
Y_{\ell}^{m} (\theta, \phi)$ is according to (\ref{YlmHomPol}) a
homogeneous polynomial of degree $\ell$ in the Cartesian components of
$\bm{r}$. Thus, it makes sense to define the differential operator
${\mathcal{Y}}_{\ell}^{m} (\nabla)$ via (\ref{Def:YlmNabla}), i.e., by
replacing the Cartesian components of $\bm{r}$ in (\ref{YlmHomPol}) by
the Cartesian components of $\nabla$.

The spherical tensor gradient operator ${\mathcal{Y}}_{\ell}^{m}
(\nabla)$ is an irreducible spherical tensor of rank $\ell$. This is a
very consequential fact. Firstly, its application to a scalar function,
which is an irreducible spherical tensor of rank $0$, must produce an
irreducible spherical tensor of rank $\ell$. This follows at once from
the simplified version (\ref{YlmNabla_phi}) of Hobson's theorem which is
discussed in Section \ref{Sec:HobsonDiffTheor}. With the help of
Hobson's theorem, it is also possible to derive the compact explicit
expression (\ref{YlmNabla_PHIlm}) for the product
${\mathcal{Y}}_{\ell_1}^{m_1} (\nabla) F_{\ell_2}^{m_2} (\bm{r})$, where
$F_{\ell_2}^{m_2}$ is an irreducible spherical tensor with nonzero rank
$\ell_2$ of the type of (\ref{Def_IrrSphericalTensor}) that also
satisfies (\ref{YlmNabla_phi}). Secondly, the structure of products of
the type of ${\mathcal{Y}}_{\ell_1}^{m_1} (\nabla) F_{\ell_2}^{m_2}
(\bm{r})$ according to (\ref{YlmNab2Flm_GenStruc}) can be understood
completely on the basis of the usual coupling rules of angular momentum
theory. Accordingly, the angular parts of such a product can be
expressed in terms of Gaunt coefficients and spherical harmonics.
Moreover, the radial parts of ${\mathcal{Y}}_{\ell_1}^{m_1} (\nabla)
F_{\ell_2}^{m_2} (\bm{r})$ and of $F_{\ell_2}^{m_2} (\bm{r})$ are
connected by relationships which only involve differentiations with
respect to the radial variable $r$. It is this property which makes the
spherical tensor gradient operator a practically useful analytical tool.

Fourier transformation is one of the principal techniques for the
evaluation of molecular multicenter integrals. It is also extremely
useful in connection with the spherical tensor gradient operator. Under
Fourier transformation, the differential operator
${\mathcal{Y}}_{\ell_1}^{m_2} (\nabla)$ produces a regular solid
harmonic ${\mathcal{Y}}_{\ell_1}^{m_2} (\mathrm{i} \bm{p})$ in momentum
space. In this way, we can easily understand the tensorial nature of
${\mathcal{Y}}_{\ell_1}^{m_2} (\nabla)$. Moreover, many analytical
manipulations can be done more conveniently in the momentum than in the
coordinate representation.

There are some scalar functions of physical interest which produce
particularly compact results if ${\mathcal{Y}}_{\ell_1}^{m_2} (\nabla)$
is applied to them. Examples are the Coulomb potential, which produces
an irregular solid harmonic according to (\ref{YllNabla_Coul}), or the
$1s$ Gaussian function, which produces a spherical Gaussian function
according to (\ref{YlmNabla_GTO}). Another class of scalar functions, to
which ${\mathcal{Y}}_{\ell_1}^{m_1} (\nabla)$ can be applied with
remarkable ease, are the so-called reduced Bessel functions defined in
(\ref{Def:RBF}), and their anisotropic generalizations, the so-called
$B$ functions defined in (\ref{Def:B_Fun}). These functions, which have
played a major role in my own research, are discussed in Section
\ref{Sec:RBF}. Application of the spherical tensor gradient operator to
a scalar $B$ function simply produces a nonscalar $B$ function according
to (\ref{STGO_Bn00}), and the product $\mathcal{Y}_{\ell_1}^{m_1}
(\nabla) B_{n_2,\ell_2}^{m_2} (\alpha, \bm{r})$ can according to
(\ref{STGO_Bnlm}) be expressed by a simple linear combination of $B$
functions. The simplicity of both (\ref{STGO_Bn00}) and
(\ref{STGO_Bnlm}) follows directly from the very simple Fourier
transform (\ref{FT_B_Fun}) of $B$ functions, which gives them an
exceptional position among exponentially decaying basis functions.

Classically, the domain of the spherical tensor gradient operator
consists of the differentiable functions $f : \mathbb{R}^3 \to
\mathbb{C}$. However, as discussed in Section \ref{Sec:SpherDeltaFun},
it makes sense to apply $\mathcal{Y}_{\ell}^{m} (\nabla)$ also to
distribution or generalized functions. This produces mathematical
objects like the spherical delta function $\delta_{\ell}^{m} (\bm{r})$,
which is defined by (\ref{Def_delta_lm}) and which can be viewed as a
generalized solution of the Poisson equation (\ref{PoissonEq_Zlm}) of a
unit multipole charge. With the help of Fourier transformation, the
nature of the spherical delta function can be made transparent, and we
can also understand easily why the Poisson equation (\ref{PoissonEq_CP})
holds, or -- to put it differently -- why convolution with the Coulomb
potential and application of the Laplace operator are inverse
operations. An essentially identical approach works also in the case of
$B$ functions. It follows from (\ref{Shift_n_Blm}) or also from the
functional equations (\ref{FuncEqs_FT_Bnlm}) that the differential
operator $1 - \alpha^{-2} \nabla^2$ of the modified Helmholtz equation
functions can be viewed as a kind of lowering operator for $B$
functions. Since the Yukawa potential $\exp (-\alpha r)/r$ is according
to (\ref{YukawaPot_Bfun}) proportional to the $B$ function $B_{0, 0}^{0}
(\alpha, \bm{r})$, it turns out that convolution with the Yukawa
potential and the application of the differential operator $1 -
\alpha^{-2} \nabla^2$ are according to (\ref{delta_B000}) inverse
operations.

In many subfields of physics and physical chemistry, an essential step
towards a solution of the problem under consideration consists in a
separation of variables. Addition theorems, which are discussed in
Section \ref{Sec:AdditionThm}, are principal tools to accomplish such a
separation of variables. As is well known, addition theorems can be
obtained according to (\ref{ExpDifOp}) by applying the translation
operator $\mathrm{e}^{\bm{r}' \cdot \nabla}$ to a function $f (\bm{r})$.
However, in atomic and molecular electronic structure calculations, we
are predominantly interested in irreducible spherical tensors of the
type of (\ref{Def_IrrSphericalTensor}), and the convenient
orthonormality of the spherical harmonics makes it highly desirable that
the functions, which occur in the addition theorem, are also irreducible
spherical tensors of the type of (\ref{Def_IrrSphericalTensor}).
Accordingly, the translation operator in its Cartesian form
(\ref{CartTransOp}) is not suited for our needs, since its use would
lead to enormous technical problems. It is a much better idea to
express the translation operator in terms of irreducible spherical
tensors according to (\ref{ST_TransOp}). The feasibility of this
approach is demonstrated by deriving the addition theorem of the
function $r^{\nu} {\mathcal{Y}_{\ell}^{m}} (\bm{r})$ with $\nu \in
\mathbb{R}$.

Let me conclude this article by some personal remarks. There can be no
doubt that $\mathcal{Y}_{\ell}^{m} (\nabla)$ is a useful analytical tool
since its application to a scalar function produces an irreducible
spherical tensor of rank $\ell$. This is a highly advantageous feature,
because now it is in principle sufficient to consider only multicenter
integrals of scalar functions. Higher angular momentum states can be
generated by differentiation with respect to the nuclear coordinates.
Often, this is simpler that the direct derivation of explicit
expressions for integrals of nonscalar functions. This approach has the
additional advantage that it facilitates the use of computer algebra
systems like Maple or Mathematica, whose systematic utilization I
wholeheartedly recommend. The completely symbolic treatment of
complicated multicenter integrals is still too difficult for computer
algebra systems, but they are already now very well suited for doing
complicated differentiations.

Nevertheless, one should never forget that a nice explicit expression
for a multicenter integral does not necessarily permit its efficient and
reliable evaluation. Already during the process of deriving an explicit
expression, one should always take numerical aspects into account,
because they ultimately decide whether a given formula is practically
useful or not. Unfortunately, this is more easily said than done. Thus,
the ability of skillfully manipulating complicated mathematical
expressions does not guarantee success in the integral business. It is
also necessary to have a detailed knowledge of numerical analysis.

For example, during the work for my PhD thesis \cite{Weniger/1982},
series expansions for multicenter integrals played a major role.
Unfortunately, it often happened that series expansions converged quite
slowly (see for example \cite[Table II]{Weniger/Steinborn/1983b}). In
principle, it is a fairly obvious idea to try to speed up the
convergence of these expansions with the help of series transformations,
but during my PhD thesis I only knew the linear series transformations
described in the classic book by Knopp \cite{Knopp/1964}, which did not
accomplish much. At that time, I was completely ignorant about the more
modern and more effective computational tools such as Pad\'{e}
approximants or other nonlinear transformations, which often accomplish
spectacular improvements of convergence.

The situation changed radically when I did postdoctoral work with
Professor Ji\v{r}{\'\i} \v{C}{\'\i}\v{z}ek at the Department of Applied
Mathematics of the University of Waterloo. There, I worked on
distributive expansions of a plane wave \cite{Weniger/1985}, which
converge weakly with respect to the norm of the Hilbert space $L^2
(\mathbb{R}^3)$ or the Sobolev space $W_2^{(1)} (\mathbb{R}^3)$. My
second research topic in Waterloo -- the summation of factorially
divergent power series as they for instance occur as asymptotic
expansions for special functions or in perturbation expansions of
quantum physics -- was in the long run far more consequential, although
I did not accomplish anything worth publishing during my stay in 1983.

After my return to Regensburg, I tried to apply nonlinear
transformations also to slowly convergent series expansions for
multicenter integrals. In some cases, remarkable improvements of
convergence were observed
\cite{Grotendorst/Weniger/Steinborn/1986,Steinborn/Weniger/1990,%
Weniger/Grotendorst/Steinborn/1986a,Weniger/Steinborn/1987,%
Weniger/Steinborn/1988}.

In order to understand better the power as well as the limitations of
nonlinear sequence transformations, I also worked on their theoretical
properties. As a by-product, I was able to derive several new
transformations. The majority of these transformations was published in
my long article \cite{Weniger/1989}, where also efficient algorithms for
the computation of sequence transformations are discussed as well as
theoretical error estimates and convergence properties. More recent
activities are described in the articles
\cite{Bender/Weniger/2001,Weniger/2000b,Weniger/2003,Weniger/2004a,%
Weniger/Kirtman/2003}.

Thus, it is probably justified to claim that numerical mathematics
ultimately profited via cross fertilizations from the convergence
problems which I encountered. My personal experience highlights the
central importance of a functioning communication between mathematicians
and scientists. Generalists like Professor Josef Paldus, who is at home
both in mathematics and in physical and theoretical chemistry, help to
overcome the current communication problems and thus do an invaluable
service to the scientific community.

\begin{appendix}
\typeout{==> Appendix A: Terminology and Definitions}
\section{Appendix: Terminology and Definitions}
\label{App:Terminolgy}

For the set of \emph{positive} and \emph{negative} integers, I write
$\mathbb{Z} = \{ 0, \pm 1, \pm 2, \ldots \}$, for the set of
\emph{positive} integers, I write $\mathbb{N} = \{ 1, 2, 3, \ldots \}$,
and for the set of \emph{non-negative} integers, I write $\mathbb{N}_0 =
\{ 0, 1, 2, \ldots \}$. The real and complex numbers and the set of
three-dimensional vectors with real components are denoted by
$\mathbb{R}$, $\mathbb{C}$, and $\mathbb{R}^3$, respectively.

For the commonly occurring special functions of mathematical physics I
use the notation of Magnus, Oberhettinger, and Soni
\cite{Magnus/Oberhettinger/Soni/1966} unless explicitly stated
otherwise.

Double factorials are defined according to
\begin{subequations}
\begin{align}
(2n)!! & \; = \; 2 \cdot 4 \cdots 2n \, , \qquad n \in \mathbb{N} \, , 
\\
(2n-1)!! & \; = \; 1 \cdot 3 \cdots (2n-1) \, , 
\qquad n \in \mathbb{N} \, ,
\\
0!! & \; = \; 1!! \; = \; (-1)!! \; = \; 1 \, .
\end{align}
\end{subequations}

Fourier transformation is used used in its symmetrical form, i.e., a
function $f: \mathbb{R}^3 \to \mathbb{C}$ and its Fourier transform
$\bar{f}$ are connected by the integrals
\begin{align}
  \label{Def_FT}
  \bar{f} (\bm{p}) & \; = \; (2\pi)^{-3/2} \int \,
  \mathrm{e}^{-\mathrm{i} \bm{p} \cdot \bm{r}} \, f (\bm{r})
  \, \mathrm{d}^3 \bm{r} \, ,
  \\
  \label{Def_InvFT}
  f (\bm{r}) & \; = \; (2\pi)^{-3/2} \int \,
  \mathrm{e}^{\mathrm{i} \bm{r} \cdot \bm{p}} \, \bar{f} (\bm{p}) 
  \, \mathrm{d}^3 \bm{p} \, ,
\end{align}

\typeout{==> Appendix B: Spherical Harmonics}
\section{Appendix: Spherical Harmonics}
\label{App:SpherHar}

Normally, the spherical harmonics $Y_{\ell}^{m} (\theta, \phi)$ are
introduced as the simultaneous normalized eigenfunctions of the orbital
angular momentum operators
\begin{equation}
  \label{Def_L^2}
\hat{\bm{L}}^2 \; = \; 
- \frac{1}{\sin (\theta)} \, \frac{\partial}{\partial \theta} \,
\sin (\theta) \, \frac{\partial}{\partial \theta} \, - \, 
\frac{1}{\sin^2 (\theta)} \, \frac{\partial^2}{\partial \phi^2} \, ,
\end{equation}
which is essentially the angular part of the three-dimensional Laplacian
in spherical polar coordinates,
\begin{equation}
  \label{Def_Lapl_r_theta_phi}
\nabla^2 \; = \; \frac{1}{r^2} \, \frac{\partial}{\partial r} \,
r^2 \, \frac{\partial}{\partial r} \, - \, \frac{\hat{\bm{L}}^2}{r^2} \, ,
\end{equation}
and 
\begin{equation}
  \label{Def_L_z}
\hat{L}_z \; = \; \mathrm{i} \frac{\partial}{\partial \phi} \, .
\end{equation}
This determines $Y_{\ell}^{m} (\theta, \phi)$ up to an arbitrary phase.
If we for instance choose the phase convention of Condon and Shortley
\cite[Chapter III.4]{Condon/Shortley/1970}, we obtain the following
explicit expression \cite[p.\ 69]{Biedenharn/Louck/1981a}:
\begin{equation}
  \label{Def_Ylm}
Y_{\ell}^{m} (\theta, \phi) \; = \; \mathrm{i}^{m + \vert m \vert} \,
\left[ \frac{(2 \ell + 1)(\ell - \vert m \vert)!}
{4 \pi (\ell + \vert m \vert)!} \right]^{1/2} \,
P_{\ell}^{\vert m \vert} (\cos \theta) \,
{\mathrm{e}}^{\mathrm{i} m \phi} \, .
\end{equation}
Here, $P_{\ell}^{\vert m \vert} (\cos \theta)$ is an associated Legendre
polynomial \cite[p.\ 155]{Condon/Odabasi/1980}:
\begin{equation}
P_{\ell}^{m} (x) \; = \; (1-x^2)^{m/2} \,
\frac{{\mathrm{d}}^{\ell+m}}{{\mathrm{d}x}^{\ell+m}} \,
\frac{(x^2-1)^{\ell}}{2^{\ell} {\ell}^{\ell}}
\; = \; (1-x^2)^{m/2} \,
\frac{{\mathrm{d}}^{m}}{{\mathrm{d}x}^{m}} \, P_{\ell} (x) \, .
\end{equation}
Alternative phase conventions for the spherical harmonics are discussed
in \cite[pp. 17 - 22]{Steinborn/Ruedenberg/1973}.

The spherical harmonics $Y_{\ell}^{m} (\theta, \phi)$ are often called
surface harmonics because the angles $\theta$ and $\phi$ characterize a
point $\bm{r}/r$ on the surface of the three-dimensional unit sphere. In
the literature, it is common to introduce the so-called regular and
irregular solid harmonics
\begin{align}
  \label{Def_RegSolHar}
  {\mathcal{Y}}_{\ell}^{m} (\bm{r}) & \; = \;
  r^{\ell} Y_{\ell}^{m} (\theta, \phi) \, ,
  \\
  \label{Def_IrregSolHar}
  {\mathcal{Z}}_{\ell}^{m} (\bm{r}) & \; = \;
  r^{-\ell-1} Y_{\ell}^{m} (\theta, \phi) \, .
\end{align}
Both ${\mathcal{Y}}_{\ell}^{m}$ and ${\mathcal{Z}}_{\ell}^{m}$ are
defined for arbitrary vectors $\bm{r} \in \mathbb{R}^3$. Moreover, it is
easy to show that the regular solid harmonics are for all $\bm{r} \in
\mathbb{R}^3$ solutions of the homogeneous three-dimensional Laplace
equation
\begin{equation}
  \label{HomogLaplaceEqn}
\nabla^2 \, f (\bm{r}) \; = \; 
\left[ \frac{\partial^2}{\partial x^2} + 
\frac{\partial^2}{\partial y^2} + 
\frac{\partial^2}{\partial z^2} \right] \, f (\bm{r}) \; = \; 0 \, ,
\end{equation}
whereas the irregular solid harmonics are solutions for all $\bm{r} \in
\mathbb{R}^3 \setminus \{\bm{0}\}$.

In connection with the differential operator ${\mathcal{Y}}_{\ell}^{m}
(\nabla)$, it is more natural to introduce the regular solid harmonics
as polynomials satisfying the Laplace equation (\ref{HomogLaplaceEqn}).
Suitable subsets of the polynomials in $x$, $y$, and $z$ can be
characterized by their transformation properties. For example, the
\emph{homogeneous} polynomials $P_{\ell} (x, y, z)$ of degree $\ell$
satisfy
\begin{equation}
P_{\ell} (\eta x, \eta y, \eta z) \; = \;
\eta^{\ell} \, P_{\ell} (x, y, z) \, , 
\qquad \eta \in \mathbb{C} \, , \quad \ell \in \mathbb{N}_0 \, .
\end{equation}
A special subset of the homogeneous polynomials of degree $\ell$ are the
so-called \emph{harmonic} polynomials $H_{\ell} (x, y, z)$ of degree
$\ell$ that are also solutions of the homogeneous Laplace equation
(\ref{HomogLaplaceEqn}), i.e., they satisfy
\begin{equation}
\nabla^2 \, H_{\ell} (x, y, z) \; = \; 0 \, .
\end{equation}
For a given $\ell \in \mathbb{N}_0$ there are exactly $2\ell+1$
\emph{linearly independent} harmonic polynomials $H_{\ell} (x, y, z)$
(see for example \cite[p.\ 123]{Hobson/1965} or \cite[Appendix
H.3]{Normand/1980}). Accordingly, the regular solid harmonics span the
space of harmonic polynomials of degree $\ell$:
\begin{equation}
H_{\ell} (x, y, z) \; = \; \sum_{m = - \ell}^{\ell} \,
C_{\ell}^{m} \, {\cal Y}_{\ell}^{m} (\bm{r}) \, .
\end{equation}

It follows from its definition (\ref{Def_RegSolHar}) that the regular
solid harmonic ${\mathcal{Y}}_{\ell}^{m} (\bm{r})$ is a homogeneous
harmonic polynomial of degree $\ell$ in the Cartesian components of
$\bm{r} = (x, y, z)$ \cite[p.\ 71]{Biedenharn/Louck/1981a}:
\begin{align}
  \label{YlmHomPol}
\mathcal{Y}_{\ell}^{m} (\bm{r}) & \; = \;
\left[ \frac{2\ell+1}{4\pi} (\ell+m)!(\ell-m)! \right]^{1/2} 
\notag \\
& \qquad \times \, \sum_{k \ge 0} \, \frac
{(-x-\mathrm{i}y)^{m+k} (x-\mathrm{i}y)^{k} z^{\ell-m-2k}}
{2^{m+2k} (m+k)! k! (\ell-m-2k)!} \, .
\end{align}

\typeout{==> Appendix C: Gaunt Coefficients}
\section{Appendix: Gaunt Coefficients}
\label{App:Gaunt}

The so-called Gaunt coefficient \cite{Gaunt/1929} is the integral of the
product of three spherical harmonics over the surface of the unit sphere
in ${\mathbb{R}}^3$:
\begin{equation}
  \label{Def_Gaunt}
\langle \ell_3 m_3 \vert \ell_2 m_2 \vert \ell_1 m_1 \rangle
\; = \; \int \, \bigl[ Y_{\ell_3}^{m_3} (\Omega) \bigr]^{*} \,
Y_{\ell_2}^{m_2} (\Omega) \,
Y_{\ell_1}^{m_1} (\Omega) \, {\mathrm{d}} \Omega \, .
\end{equation}
Gaunt coefficient can be expressed in terms of Clebsch-Gordan
coefficients \cite[Eq.\ (3.192)]{Biedenharn/Louck/1981a} or $3jm$
symbols \cite[p.\ 168]{Condon/Odabasi/1980}, respectively:
\begin{align}
  \label{Gaunt_ClebschGordan} 
  \langle \ell_3 m_3 \vert \ell_2 m_2 \vert \ell_1 m_1 \rangle 
  & \; = \; 
  \left[ \frac{(2\ell_1+1)(2\ell_2+1)} {4\pi (2\ell_3+1)}
  \right]^{1/2} \, 
  C^{\ell_1 \ell_2 \ell_3}_{0 0 0} \, 
  C^{\ell_1 \ell_2 \ell_3}_{m_1 m_2 m_3} 
\\[1.5\jot]
  \label{Gaunt_3jm}
& \; = \;
  (-1)^{m_3} \left[ \frac {(2\ell_1+1)(2\ell_2+1)(2\ell_3+1)} {4\pi}
  \right]^{1/2}
  \notag \\[1.5\jot]
  & \qquad \times \,
 \begin{pmatrix}
   \ell_1 & \ell_2 & \ell_3 \\
   0 & 0 & 0
 \end{pmatrix} \, 
 \begin{pmatrix}
   \ell_1 & \ell_2 & \ell_3 \\
   m_1 & m_2 & - m_3
 \end{pmatrix} \, .
\end{align}

It follows from the orthonormality of the spherical harmonics that the
Gaunt coefficients linearize the product of two spherical harmonics:
\begin{equation}
  \label{Ylm_lin}
Y_{\ell_1}^{m_1} (\Omega) \, Y_{\ell_2}^{m_2} (\Omega) \; = \;
\sum_{\ell=\ell_{\mathrm{min}}}^{\ell_{\mathrm{max}}}
\! {}^{(2)} \,
\langle \ell m_1+m_2 \vert \ell_1 m_1 \vert \ell_2 m_2 \rangle
Y_{\ell}^{m_1+m_2} (\Omega) \, .
\end{equation}
The symbol $\sum \! {}^{(2)}$ indicates that the summation proceeds in
steps of two. The summation limits in (\ref{Ylm_lin}), which follow from
the selection rules of the $3jm$ symbols, are given by
\cite[Eq.\ (3.1)]{Weniger/Steinborn/1982}
\begin{subequations}
  \label{SumLims}
\begin{align}
  \label{SumLims_a}
  \ell_{\mathrm{max}} & \; = \; \ell_1 + \ell_2 \, ,
  \\
  \label{SumLims_b}
  \ell_{\mathrm{min}} & \; = \;
\begin{cases}
  \lambda_{\mathrm{min}} \, , \qquad \; \, \, \text{if} \;
  \ell_{\mathrm{max}} + \lambda_{\mathrm{min}} \; \text{is even} \, ,
  \\[1.5\jot]
  \lambda_{\mathrm{min}} + 1 \, , \quad \text{if} \; \ell_{\mathrm{max}}
  + \lambda_{\mathrm{min}} \; \text{is odd} \, ,
\end{cases} \\
   \label{SumLims_c}
   \lambda_{\mathrm{min}} & \; = \; \max (\vert \ell_1 - \ell_2 \vert,
   \vert m_1 + m_2 \vert) \, .
\end{align}
\end{subequations}

Gaunt coefficients can be computed according to either
(\ref{Gaunt_ClebschGordan}) or (\ref{Gaunt_3jm}) via the numerous
explicit expressions for Clebsch-Gordan coefficients or $3jm$ symbols
described in the literature. In the case of small angular quantum
numbers, this is a satisfactory approach, but in the case of large
angular quantum numbers, the computation of individual Clebsch-Gordan
coefficients or $3jm$ symbols via their explicit expressions becomes
computationally costly. In addition, a catastrophic accumulation of
rounding errors can easily happen. A much better approach consists in
the use of a recurrence formulas for $3jm$ symbols derived by Schulten
and Gordon \cite{Schulten/Gordon/1975}. Although this recursion is
neither stable upwards nor downwards, it is nevertheless possible to
compute in this way whole strings of $3jm$ symbols efficiently and
reliably even for very large angular momentum quantum numbers
\cite{Schulten/Gordon/1976}. In \cite{Weniger/Steinborn/1982}, a FORTRAN
program is described that computes simultaneously all Gaunt coefficients
occurring on the right-hand side of (\ref{Ylm_lin}) with the help of the
recurrence formula and the computational algorithm of Schulten and
Gordon \cite{Schulten/Gordon/1975,Schulten/Gordon/1976}.

It may be of interest for the reader that several articles on Gaunt
coefficients have appeared in recent years
\cite{Homeier/Steinborn/1996b,Xu/1996,Xu/1997,Xu/1998,Sebilleau/1998,%
Mavromatis/Alassar/1999}. Then, there are some very recent articles by
Dunlad \cite{Dunlap/2002,Dunlap/2003,Dunlap/2005}, but the objects he
considered are not the usual Gaunt coefficients defined in (\ref{Def_Gaunt})
but rather generalizations that occur in the context of molecular
multicenter integrals of spherical Gaussian functions.

In this article, the following abbreviations will frequently be used:
\begin{align}
  \label{Def_Del_l}
\Delta \ell & \; = \; (\ell_1 + \ell_2 - \ell)/2 \, ,
\\
  \label{Def_Del_l_1}
\Delta \ell_1 & \; = \; (\ell - \ell_1 + \ell_2)/2 \, ,
\\
  \label{Def_Del_l_2}
\Delta \ell_2 & \; = \; (\ell + \ell_1 - \ell_2)/2 \, ,
\\
  \label{Def_sigma_l}
\sigma (\ell) & \; = \; (\ell_1 + \ell_2 + \ell)/2 \, .
\end{align}
If the three orbital angular momentum quantum numbers $\ell_1$,
$\ell_2$, and $\ell$ satisfy the summation limits (\ref{SumLims}), then
these quantities are always positive integers or zero.
\end{appendix}

%
%
%
{\small
\providecommand{\SortNoop}[1]{} \providecommand{\OneLetter}[1]{#1}
  \providecommand{\SwapArgs}[2]{#2#1}

}
%

\end{document}